\documentclass[printer]{aa}
\usepackage{graphicx}
\usepackage{bm}
\usepackage{amsmath}
\usepackage[T1]{fontenc}
\usepackage{lmodern}
\let\la=\langle
\let\ra=\rangle

\begin{document}
\title{Statistical evolution of quiet-Sun small scale magnetic features 
using Sunrise observations}
\author{L.~S.~Anusha$^{1}$, Sami K. Solanki$^{1,2}$, Johann Hirzberger$^1$, 
Alex Feller$^1$}

\institute{Max-Planck-Institut f{\" u}r Sonnensystemforschung,
Justus-von-Liebig-Weg 3, 37077 G\"ottingen, Germany
\email{bhasari@mps.mpg.de}
\and 
School of Space Research, Kyung Hee University,
Yongin, 446-701 Gyeonggi, Republic of Korea
}
\titlerunning{Evolution of small scale magnetic fields using Sunrise data}
\authorrunning{Anusha et al.}
\abstract{
The evolution of small magnetic features in quiet regions of the Sun provides a unique
window to probing solar magneto-convection.
Here we analyze small scale magnetic features in
the quiet Sun, using the high resolution, seeing-free observations from the Sunrise balloon 
borne solar observatory. Our aim is to understand the contribution of different physical 
processes, such as splitting, merging, emergence and cancellation of magnetic 
fields to the rearrangement, addition and removal of magnetic flux in the
photosphere. We employ a statistical approach for the analysis and the evolution
studies are carried out using a feature tracking technique. In this paper
we provide a detailed description of the feature tracking algorithm
that we have newly developed and we present the results of a statistical study of
several physical quantities. The results on the fractions of the flux 
in the emergence, appearance, splitting, merging, disappearance
and cancellation qualitatively agrees with other recent studies
\citep[see e.g.,][]{lambetal2008,lambetal2013}. To summarize, 
the total flux gained in unipolar appearance is an order of magnitude larger than the 
total flux gained in emergence. On the other hand, the bi-polar 
cancellation contributes nearly an equal amount to the loss of magnetic flux as 
unipolar disappearance. The total flux lost in cancellation 
is nearly $6-8$ times larger than the total flux gained in emergence.
One big difference between our study and previous similar studies 
is that thanks to the higher spatial resolution of Sunrise we can 
track features with fluxes as low as $9\times10^{14}$ Mx. 
This flux is nearly an order of magnitude lower than the smallest fluxes of the 
features tracked in the highest resolution previous studies based on Hinode data.
The area and flux of the magnetic features follow power-law type 
distribution, while the lifetimes show either power-law or exponential type 
distribution depending on the exact definitions used to define various birth 
and death events. We also statistically determine the evolution of the flux 
within the features in the course of their lifetime, finding that this 
evolution depends very strongly on the birth and death process that the 
features undergo.
}

\maketitle

\section{Introduction}
The Sun's magnetic field is to a large part concentrated into small-scale structures. This is 
particularly so in the quiet Sun. Although much is known about the structure of these 
features, there are still many unknowns \citep[see e.g.,][for a review of quiet-Sun 
magnetism]{dewijnetal2009}. Best studied are properties such as the field strength and the 
inclination angle of the field, although even for these widely different results have been 
published. The average magnetic field strength in network regions is known to be of on
the order of $1-2$ kilo-Gauss \citep[see e.g.,][]{stenflo1973,solankiandjos84,rabin92}, but it is rather 
difficult to estimate inter-network quiet-Sun magnetic fields because different methods 
and different spectral lines have often provided discrepant results. For example, on the one hand, 
investigations using infrared (e.g., $1.5$ $\mu$m) lines give generally low field strengths 
below 600 G\,\citep[see e.g.,][]{lin95,solankietal96,khomenkoetal2003,khomenkoetal2005b}, 
but with a tail of kG fields in the histogram of the field strengths shown by 
\citet[][]{khomenkoetal2003}. On the other hand, investigations using the visible 
line pairs (e.g., $525.0$ nm, $524.7$ nm, and $630.25$ nm, $630.15$ nm) have provided an 
inhomogeneous picture \citep[see e.g.,][]{socasnavarroandlites2004,sanchezalmedaetal2003,
khomenkoetal2005a,orozcoetal2007,martinezgonzalezetal2008,laggetal2010}, although 
in the recent studies the tendency has been towards a predominance of fields with low strengths 
\citep[see, e.g.][]{aaramosandmartinez2014}.
The magnetic field inclination calculations differ in various studies and are also not 
well established yet. Investigations of Stokes profiles observed with Hinode/SP by 
\citet[][]{orozcoetal2007} and \citet[][]{litesetal2008} suggest that the magnetic fields 
in internetwork regions are inclined predominantly horizontally with an average inclination angle of 
$20^{\circ}$ to the horizontal. However a study of bright points in Ca {\sc ii} H 
images observed using SUNRISE/SuFI, which have a mean area close to $0.01$ Mm$^2$ (nearly 
$100$ times smaller than a typical granule), by \citet[][]{jafarzadehetal2013} 
suggests that these magnetic elements in the internetwork regions are predominantly 
oriented vertically \citep[][]{jafarzadehetal2014b}. For a comprehensive review on 
this topic we refer to \citet[][and the references cited therein]{borreroetal2015}. 

The number of studies that have followed the evolution of individual features, 
or have followed a group of features through a certain phase of their evolution, is smaller. 
Such studies have mainly concentrated on individual phases in the lifetime of magnetic features, 
or particular dynamic processes, such as  be the emergence of small loops 
\citep[see e.g.,][]{centenoetal2007,martinezgonzalezetal2010,
guglielminoetal2012} and the linear polarization patches often associated with such emergence
\citep[][]{litesetal2008,danilovicetal2010}, the convective collapse of small-scale magnetic elements 
\citep[see e.g.,][]{nagataetal2008,fischeretal2009}, the evolution 
following the collapse \citep[see e.g.,][]{requereyetal2014,requereyetal2015,utzetal2014}, the 
evolution of the brightness \citep[see e.g.,][]{bergeretal2007,jafarzadehetal2013},  
and motions of bright points \citep[see e.g.,][]{dewijnetal2008,abramenkoetal2011,jafarzadehetal2014a}.
Useful as they are, such studies cannot give a full picture of the evolution of magnetic features, 
since they do not attempt to give a complete statistical description. 

A statistically more complete description has been attempted by a number of authors in recent 
years. Such studies have included not just the emergence, appearance, cancellation and 
disappearance of magnetic features, but have also taken into account 
the splitting and merging of the features, which are found to play an important role  
\citep[see][and the references cited therein]{deforestetal2007,lambetal2008,lambetal10,
parnelletal09,lambetal2013,iidaetal2012,gosicetal2014,zhouetal2010,zhouetal2013,iidaetal2015}.
Such investigations have turned up some surprising results. Thus, only a small fraction 
of the magnetic flux in the quiet Sun is observed to emerge as bi-polar 
structures and just as small a fraction of flux is removed through 
cancellation of opposite polarity fields \citep[see e.g.,][]{lambetal2013,iidaetal2015}. 

Such unexpected results demand a detailed analysis 
using an independent algorithm and the best data currently available. 
It is important to point out the contribution of the ground-based telescopes that provide 
high-spatial resolution data such as the NST of the Big Bear Solar Observatory 
\citep[e.g.,][]{goodeetal2010} and the Swedish Solar Telescope of La Palma 
observatory \citep[e.g.,][]{scharmeretal2003}.
However, space or balloon-borne observatories have added advantages such as 
obtaining seeing-free observations, or of providing data at UV wavelengths 
that can be used to link the photosphere with higher layers and which are not 
possible from ground.

Here we present results obtained with a newly developed code that carefully classifies 
the features that participate in events such as simple appearance and
disappearance, splitting, merging, bi-polar emergence and cancellation. 
The algorithm tracks the magnetic features by assigning a unique 
birth, unique death and a well defined lifetime to each of the detected 
magnetic feature. This code is applied to time series of 
high resolution magnetograms obtained in the quiet Sun with the Imaging Magnetograph 
eXperiment \citep[IMaX, see][]{martinezpilletetal2011} on the Sunrise balloon-borne 
telescope \citep[see][]{solankietal10,bartholetal11}. 
The earlier studies using high-resolution data to which our results can be 
compared have used Hinode/NFI data with a spatial resolution of 0\arcsec.3 
($6\times10^{15}$ Mx, $6.5\times10^{15}$ Mx and $5\times10^{16}$ Mx 
being the lowest detected values of flux per feature reported by 
\citealt{zhouetal2013}, \citealt{gosicetal2014} 
and \citealt{thorntonandparnell2011}, respectively). 
The higher spatial resolution (a factor of 5 per pixel, see Sect.~\ref{data}) 
and the good polarimetric sensitivity of Sunrise/IMaX data means that we can detect 
and follow features with considerably lower flux ($9\times10^{14}$ Mx is the 
lowest detected flux per feature) than previous investigations. 
The flux per feature is significantly larger than 
the often quoted flux per pixel because of the usually near-critical sampling of the 
diffraction limit of the telescope and the need to clearly distinguish 
features from noise. The fact that we can detect an order of magnitude lower flux
than previous studies has the benefit that
we can check if some of the seeming appearances and disappearances found in
earlier studies are real or are only apparent due to limited sensitivity.
We then carry out an analysis of the statistical properties 
of these features, such as distributions of area, magnetic field, magnetic fluxes, 
lifetimes and their relationships. 

In Sect.~\ref{data} we briefly describe the observations. Sect.~\ref{procedure}
is dedicated to a detailed description of the analysis procedure. In Sect.~\ref{results}
we present a discussion of our statistical studies.
Conclusions are drawn in Sect.~\ref{conclusions} and details of the method are 
given in the two appendices.

\section{The data}
\label{data}
High spatial and temporal resolution data recorded by the Sunrise 
observatory \citep[][]{solankietal10,bartholetal11,berkefeldetal2011,
gandorferetal2011,martinezpilletetal2011}
on its first science flight in June 2009 provide us with an 
opportunity to derive the statistical properties and follow the
evolution of quiet-Sun magnetic structures down to lower flux levels and at
higher spatial resolution than previous studies.

A time series of forty two quiet Sun magnetograms observed with IMaX 
\citep[][]{martinezpilletetal2011} was 
considered. The spatial sampling is approx. $40$ km/pixel
(with the spatial resolution being roughly $0.15-0.18$ arcsec)
and the cadence is $33$ sec. Observations are available 
at four wavelength points within the spectral line Fe {\sc i} at $525.02$ nm 
and at a wavelength in the nearby continuum $227$ m\AA\, from the line core 
(V5/6 data). Four wavelength points sample the line 
at $\pm$ $40$ and $\pm$ $80$ m\AA\, from the line center. 
We use phase diversity reconstructed data 
\citep[see e.g.,][]{gonsalves82,paxmanetal96,vargas2009} for descriptions of the 
technique and \citet[][]{martinezpilletetal2011} for its application to Sunrise/IMaX data.

\begin{figure*}
\begin{center}
\sidecaption
\includegraphics[width=12cm]{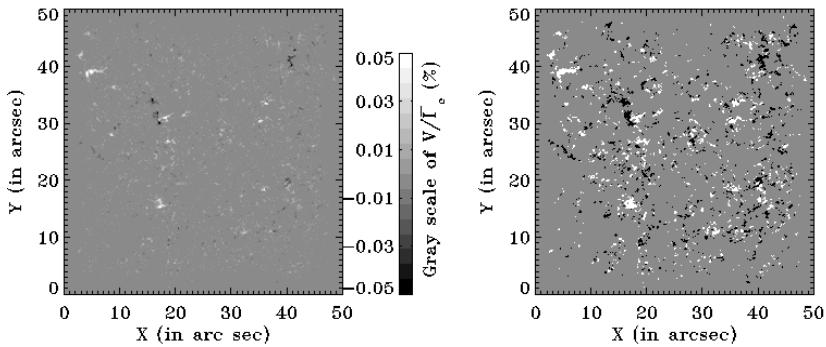}
\caption{Left: Original $V/\bar{I}_c$ image at the middle of the time-series.
Right: After applying the binary mask to the image on the left, as described in the
main text.}
\end{center}
\label{fig-data}
\end{figure*}

We then consider the 
pixels in this set of images as the points in a cuboid. Two spatial 
co-ordinates ($x,y$) and a time co-ordinate ($t$) form the three dimensions of the 
cuboid. Application of phase diversity reconstruction to the data leads to the 
loss of $108$ pixels \citep[][]{martinezpilletetal2011}. 
The features in the first and the last time steps are identified and tracked but not used
for the statistics to avoid boundary errors in feature identification. The
noise level of a single reconstructed IMaX image is $3\times 10^{-3}$ $I_c$, where 
$I_c$ is the continuum intensity. The effective size of 
usable image is $\sim 43 \arcsec \times 43 \arcsec$. Due to the low signal levels in 
Stokes $Q$ and $U$ \citep[][]{danilovicetal2010} we
concentrate on Stokes $I$ and $V$ in this paper. 

At each time step, the spatially averaged continuum intensity $\bar{I}_c$ is used
for the normalization of Stokes parameters $V$ and $I$ to obtain $V/\bar{I}_c$ and 
$I/\bar{I}_c$ at each wavelength. 
Then, the $V/\bar{I}_c$ data are averaged over the four wavelengths after reversing
the sign of Stokes $V$ in the red wing  of the line. 
We have excluded the continuum point for this averaging. We note here that we 
have used the wavelength averaged $V/\bar{I}_c$ only to identify the features as
described in Sect.~\ref{procedure}. This step does not affect the 
LOS magnetic field computation discussed in Sect.~\ref{flux-computation}. The latter uses
$V({\lambda})$ and $I(\lambda)$ individually at all wavelength points.

\section{The analysis procedure}
\label{procedure}

\subsection{Identifying the magnetic features}
A feature identification and tracking algorithm is necessary for studies of 
the evolution of magnetic features. 
A number of feature identification and tracking algorithms have been
developed by several groups, a summary of which can be found in
\citet[][]{deforestetal2007}. 

In this paper, a lane finding code originally developed for identifying granular structures 
\citep[][]{hirzbergeretal99} is used to find all magnetic patches satisfying a set of 
predetermined criterion. This code has been extensively tested and applied
to the study of supergranules by \citet[][]{hirzbergeretal08}.
The patches identified in this manner are then referred to as 
``features''. A feature is defined to be a collection of spatially connected pixels 
whose Stokes $V/I_c$ signal lies above 
a threshold of $3\times 10^{-3}$. This corresponds to two times the noise level, 
$\sigma$, of the data after the averaging over the 4 wavelength pixels in the line. 

Since $2\sigma$ is a relatively low threshold, we, in addition, remove all 
features with an area less than 5 pixels. 
Further, the features touching the spatial boundaries at any time in the course of their evolution 
are also removed, since we only have a lower limit of their area and may be missing interactions with
features outside the frame. We identify positive and negative polarity features separately. 
After the various steps described above, we obtain a total of 50255 features
of both polarities, which were considered for the analysis. 
In Fig.~\ref{fig-data} we display $V/\bar{I}_c$ image at the middle of the time 
series before (left panel) and after (right panel) identifying the features. 
Since we remove the features touching the spatial boundaries, we do not see features near 
the edge of the field-of-view.
The code also checks if the features that are originally 
identified individually in different snapshots are the same underlying magnetic feature 
seen at two points in time. 
After that the binary masks of two consecutive 
time steps are compared with each other to see if a feature present in one mask is 
still present in the subsequent mask. In general, if two individually identified features in 
two consecutive time steps share at least one spatial pixel, then they are identified 
as the same feature at two different times. 
In practice, a number of other possibilities also occur. These are described in 
detail in Sect.~\ref{def}.

\subsection{Flux Computation}
\label{flux-computation}
\begin{figure}
\centering
\includegraphics[scale=0.45]{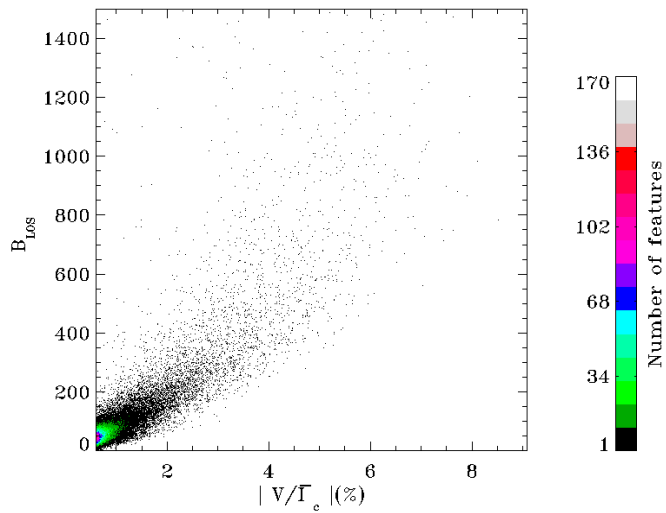}
\caption{
Magnetic flux computed using 
the COG technique. 
The figure shows a 2D histogram of the number of features on the 
plane spanned by $B_{\rm{LOS}}$ and $|V/\bar{I}_c|$ for each pixel.
}
\label{fig-flux-calib1}
\end{figure}

Using the center of gravity (COG) technique \citep[see][]{reesandsemel1979}, the 
strength of the line-of-sight (LOS) component of the magnetic field vector is determined from Stokes $V$ 
through the relationship
\begin{equation}
B_{\rm{LOS}}=|\Delta \lambda_G/ C_0 g \lambda_0^2|,
\label{cog1}
\end{equation}
where $C_0=4.67\times10^{-13}$m$^{-1}$\,G$^{-1}$, $g$ is the Land\'e factor, $\lambda_0$
is the line center wavelength, and
\begin{equation}
\Delta \lambda_G = 
\frac{\int_{-\infty}^{+\infty}{V}\Delta \lambda d\Delta \lambda }
{\int_{-\infty}^{+\infty}(I_c-I) \,\,d\Delta \lambda}.
\label{cog2}
\end{equation}
Here $I$ and $V$ represent the intensity and the circular polarization at a 
given wavelength, respectively, and $I_c$ the corresponding continuum intensity.
The LOS magnetic field values $B_{\rm{LOS}}$ 
obtained when applying this technique to each pixel range between 0 to 2000 G. 
A SIR inversion sampling a region of the same data also provided
kG fields \citep[see][]{laggetal2010,requereyetal2014}.
We note here that these magnetic field values determined using the COG technique are
averaged over the SUNRISE/IMaX spatial resolution element (i.e. no filling factor is 
introduced). A two-dimensional (2D) histogram of the number of features 
on the plane spanned by $B_{\rm{LOS}}$ and
$|V/\bar{I}_c|$ is shown in Fig.~\ref{fig-flux-calib1}. 
We use Equations~(\ref{cog1} and \ref{cog2}) for computing the flux. 
However, since it is useful to show $|V/\bar{I}_c|$ on the horizontal axis of 
Fig.~\ref{fig-flux-calib1}, we can trivially divide both numerator and denominator of 
Equation~(\ref{cog2}) by $\bar{I}_c$ (which is the continuum 
intensity averaged spatially over the entire image for each time frame). 
This step does not affect $B_{\rm{LOS}}$ values.
It is cut-off at 1450 G due to the small number of pixels with large $B_{\rm{LOS}}$.
Besides a significant scatter, the plot displays also a non-linear trend that
is partly due to Zeeman saturation \citep[][]{stenflo1973} and partly to
temperature weakening of the spectral line \citep[][]{zayeretal1990}, although the latter 
effect is not as extreme as found in MHD simulations by \citet[][]{shelyagetal2007}.
Due to the influence of noise the $B_{\rm{LOS}}$ thus obtained cannot be trusted in pixels
with low signal. Therefore in the following we determine and use a feature-averaged value, 
$\la B_{\rm{LOS}} \ra$, for computing the magnetic flux of a feature. The quantity 
$\la B_{\rm{LOS}} \ra$ is determined by averaging the Stokes $V$ data as discussed in the 
following paragraph. Here the angled brackets $\la\,\,\,\ra$ represent the 
value obtained from the spatial average of Stokes $V$ data over individual 
magnetic features.

\begin{figure}
\centering
\includegraphics[scale=0.45]{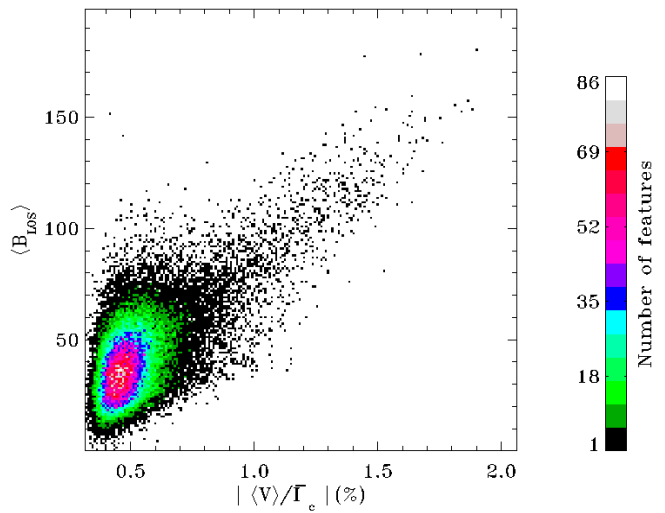}
\caption{
Magnetic flux computed using 
the COG technique. A 2D histogram of the number of features
on the plane spanned by the magnetic feature averaged LOS component of the field 
$\la B_{\rm{LOS}} \ra$ and the feature averaged unsigned Stokes $V$ normalized to the
continuum intensity namely, $|\la V/\bar{I}_c \ra|$.
}
\label{fig-flux-calib2}
\end{figure}

Thus, we compute the $\la B_{\rm{LOS}} \ra$ (Equations~\ref{cog1} and \ref{cog2})
using spatially averaged quantities $\la V(\lambda) \ra$, $\la I(\lambda) \ra$ and $\la I_c(\lambda) \ra$
within each of the 50255 identified magnetic features. A 2D histogram of the number of features
on the plane spanned by $\la B_{\rm{LOS}} \ra$ and
$|\la V \ra/\bar{I}_c|$ is shown in Fig.~\ref{fig-flux-calib2}.
The obtained averaged field values range between 0 and 100 G. Only 410 features 
out of 50255 features have field strengths between 100 and 200 G. The strong 
reduction in $B_{\rm{LOS}}$ values is caused by the inclusion of the weak 
signals at the edges of the magnetic features
in the averaging. A higher threshold than the $2 \sigma$ value that we have imposed leads
to higher average $B_{\rm{LOS}}$ values as test calculations showed. However, since many 
features are then lost, we decided to maintain the lower threshold value. 
The thus obtained $\la B_{\rm{LOS}} \ra$ is then multiplied 
by the area covered by the feature to obtain its magnetic flux. 

\subsection{Definitions of features, parameters and events}
\label{def}
\begin{figure}
\begin{center}
\includegraphics[scale=0.45]{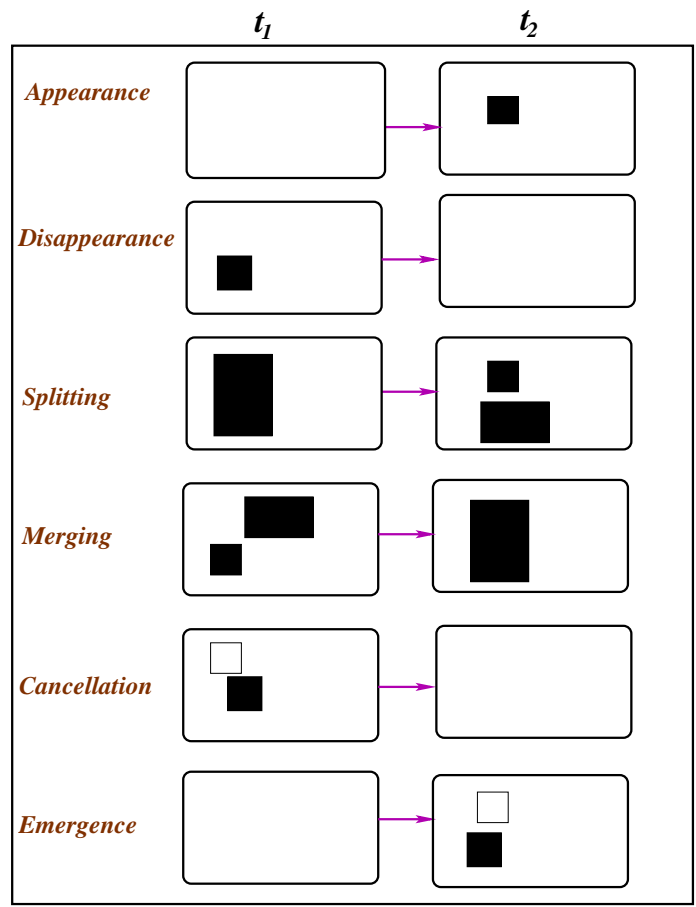}
\end{center}
\caption{Illustration of various possible interactions between the magnetic features. 
Only the simplest possibilities are shown in this diagram. 
Black and white refer to magnetic features with opposite polarities from each other.
A more detailed description of all possible events is provided in the main text.}
\label{fig-events}
\end{figure}

In this section we define and describe different terms, quantities and
events on which our statistical study in this paper is based. 
The conventions defined here are used throughout this paper. 
Fig.~\ref{fig-events} illustrates various possible interactions between the
features, that are defined in this section.

Let us consider two successive time steps $t_1$ and $t_2=t_1+\Delta t$,
where $\Delta t$ is the time step. From $t_1$ to $t_2$ important properties 
of a given feature that can change are, (1) spatial location of the feature, 
(2) shape and area of the feature, and (3) magnetic flux contained in 
that feature. In addition, the feature can disappear, or cancel with an 
opposite polarity feature, or merge with several same-polarity features, or 
split into multiple features between the times $t_1$ and $t_2$. Finally, the feature 
can appear or emerge (together with an opposite polarity feature) between 
the times $t_1$ and $t_2$. To quantitatively describe such evolutionary steps, 
several properties and events need to be defined. \\

\noindent
$\bullet$ ``Size'' or ``area'' of a feature denoted as $A$, is given by the 
total number of pixels covered by a feature. \\

\noindent
$\bullet$ Birth of a feature by appearance: 
If a feature appears at time $t_2$ at a given spatial location, but if
the same spatial location at $t_1$ was empty (i.e. at $t_1$ no magnetic feature
overlapped with any of the pixels occupied by the features at $t_2$), 
then we say that a feature is born 
by appearance at $t_2$. \\

\noindent
$\bullet$ Death of a feature by disappearance: If a feature is present at $t_1$ 
at a given spatial location, but if the same spatial location 
at $t_2$ is found to be empty (i.e. at $t_2$ no magnetic feature
overlaps with any of the pixels occupied by the features at $t_1$), 
then we say that the feature died or disappeared 
at $t_1$. \\

\begin{figure*}
\begin{center}
\sidecaption
\includegraphics[width=12cm]{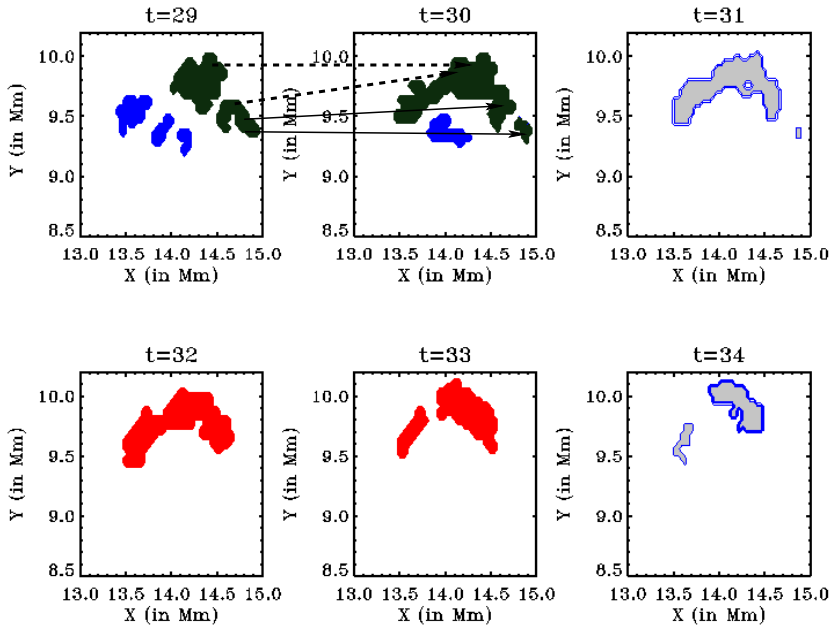}
\caption{Illustrative example of evolution of a group of magnetic features over 6 time steps. 
All magnetic features in this frame have the same polarity. Features involved in merging and 
splitting events are indicated by filled colored contours. 
The arrows in the first two panels refer to a particular event involving 
simultaneous splitting and merging, as discussed in the main text.
The dashed arrows indicate the merging event while solid line arrows
mark the splitting event. 
The features that are involved in this event in the first 2 panels 
are colored blue and green. The features that do not 
split or merge in the plotted frame are shaded gray.  
The order of time steps proceeds row by row, and from left to right 
in each row. 
}
\label{fig-668-arbitrary}
\end{center}
\end{figure*}

\noindent
$\bullet$ Splitting of a feature: A feature at $t_1$ is said to split, if it breaks into 
two or more features at $t_2$. This is taken to happen when the features at $t_2$  
(the children) have the same polarity as the parent, are spatially separated 
from each other and overlap with the parent by at least one pixel each. 
For the purposes of this paper we say that the parent feature died at $t_1$ due to splitting. 
At the same time, all the children at $t_2$ are considered to be newly born. 
An exception is made when the overwhelming part of the area or 
magnetic flux of the parent lives on in just one child, 
i.e. when only a small piece breaks off a much larger feature. Then the parent is assumed
to live on as the biggest of the children. All other children are assumed to be newly born.
This is described more quantitatively under `area-ratio criterion' below.
In Fig.~\ref{fig-668-arbitrary} examples of splitting events are 
shown in red. \\

\noindent
$\bullet$ Merging of features : Merging is said to have taken place if two or more
features of the same polarity at $t_1$ combine into a single feature of the same polarity 
at $t_2$, and the child overlaps spatially by at least one pixel with each of the 
parent features. For the purposes of this paper we say that all parent features died at $t_1$ 
due to merging and a new feature is born at $t_2$. Again, an exception is made if one
parent is overwhelmingly larger than the other(s).
In Fig.~\ref{fig-668-arbitrary} 
examples of merging events are shown in blue.\\

\noindent
$\bullet$ To rule out that a feature is pronounced dead just because a tiny part of 
it breaks off, or because it merges with a much smaller feature, the areas of 
involved features at a given time step are compared.
We compare areas of child features for a splitting event and areas of 
parent features for a merging event. 
If the largest feature is more than a predetermined factor times bigger than the
second largest, then this feature is assumed to be a continuation of the parent (in the 
case of splitting) or to continue as the child (in the case of merging).
All other children are classified to be born due to splitting-off birth events.
To test how strong the effect of the above area-ratio is, we consider 
the following set of area-ratios: 2:1, 3:1, 5:1 and 10:1. 
For example, in a 2:1 area ratio criterion for a splitting event, 
the parent is considered dead only if the area of the largest child 
is less than twice the area of second largest child. 
A quantitative description of
splitting and merging events is given in Appendix~\ref{appendixa}.

The area-ratio criterion also helps to bring order to complex events, where splitting and 
merging may be taking place at the same time. An illustration of this is 
shown in Fig.~\ref{fig-668-arbitrary}. 
Let us focus on the first two panels (first row from left to right). The four features 
involved in a complex event are the filled green features connected by 
arrows (see figure caption).
In principle, since there are two features at both time steps one could say that no 
interaction has taken place. However, one of these features has grown considerably, 
while the other has shrunk. In addition, one of the features at $t_1=29$ overlaps with 
two features at $t_2=30$ and vice versa. Such a situation could be the result of a 
combination of a splitting event plus a merging, both taking place within a single 
time-step, but it can also be produced by other combinations of events. The chosen 
area-ratio criterion along with a set of priority criteria described in 
Appendix~\ref{appendixb} to eliminate the ambiguity (see next paragraph) leads to one 
such combination being preferred. After applying an area-ratio criterion of 5:1 
and the priority criterion to eliminate ambiguity the splitting event is discarded, 
the merging event is retained, and the remaining small green feature in the 
second panel is considered to be newly born by a splitting-off birth event.
A blue filled feature in panel 2 of Fig.~\ref{fig-668-arbitrary} dies 
by disappearance and the other two filled green features continue to live as 
gray filled features in panel 3. The smaller gray filled feature in panel 3 disappears 
and the larger gray feature continues to live as the only red filled feature in panel 4.
This eventually splits into two red filled features in panel 5, both of which continue to
live as gray filled features in panel 6.\\

\noindent
$\bullet$
Simultaneous events and ambiguities: 
Statistically, it is not so rare for splitting and merging events to 
happen simultaneously. An example of a possible simultaneous splitting and merging
event was discussed in the previous paragraph (filled green features in first two panels of 
Fig.~\ref{fig-668-arbitrary}). The events are marked by arrows (see figure caption). 
Therefore, additional care must be taken to handle such 
cases. In Appendix~\ref{appendixb} we describe in detail all possible 
simultaneous events and how we handle them. Eventually based on a set of rules all
the ambiguities in simultaneous events are eliminated and are 
classified into different birth and death cases defined above.\\

\noindent 
$\bullet$
The lifetime of a feature is defined to be the interval of time between 
the birth of a feature (either through appearance, emergence, or from splitting 
or merging) and its death (through disappearance, cancellation, splitting or 
merging).  \\

\noindent
$\bullet$
Emergence events are composed of the appearance of opposite polarity
features close in space and time. First, a spatial overlap (after dilation, 
see Appendix~\ref{appendixc}) of one or more positive
features with one or more negative features are detected. Then we apply a 
flux-ratio criterion between the sets of positive and negative features.
We consider four cases, namely, 10:1, 5:1, 3:1 and 2:1. If the ratio of sum 
of the fluxes of the interacting positive polarity features to that in the 
interacting negative polarity features satisfies the imposed flux-ratio 
criterion, then it is identified as an emergence event. We describe what we mean by 
``close in space and time'' in Appendix~\ref{appendixa}.
We classify the emergence into two types: (a) Time-symmetric
and (b) Time-asymmetric bi-polar emergence.\\

\noindent
(a) Time-symmetric emergence: The spatially close (neighboring) opposite polarity
features appear at the same time step. In this case, the total flux contained in the 
emergence event is $F_{\rm {SYM,EM}}=2\,\, {\textrm{min}}(F_+,F_-)$,
where $F_+$ is the total flux of the set of positive features emerged and, $F_-$, that
of the negative features emerged.\\

\noindent
(b) Time-asymmetric emergence: The opposite polarity features appear spatially close to each 
other at a given time-step, but the time frames at which the opposite polarity features 
are born, are offset by one or more steps. Here, only features of one of the polarities
are newly visible near a set of close-by opposite polarity features that already exist. 
The already existing features could have been born in any of the previous time steps. 
In this case, the total flux contained in the emergence event is 
$F_{\rm {ASYM,EM}}=2\,\,F_+$ if the newly emerged features are positive 
and $F_{\rm {ASYM,EM}}=2\,\, F_-$ if the newly emerged 
features are negative. We have implicitly assumed that a bi-polar feature has emerged, 
but that one polarity appears within a patch of pre-existing field. \\ 

We note here that although we have defined time-asymmetric emergence events
such that the time-separation between the already existing features and newly emerged features
is one or more, for most of the cases, the time-separation is small. 
For e.g. for the 10-1 area-ratio criteria, the maximum time-separation between the opposite polarity
features in time-asymmetric emergence is $\sim$ 9 minutes. However, the number of features having
a time-separation greater than 5 minutes is only 4 and that greater than 2 minutes is only 10 \%
of the total number of features appearing as already existing time-asymmetric counterparts.
 \\

\noindent
$\bullet$
Cancellation events occur when two or more neighboring opposite polarity 
features at time step $t_1$ exhibit a spatial overlap (after dilation, 
see Appendix~\ref{appendixc}) in such a way that 
in the subsequent time step $t_2$ the sum of the fluxes of these features 
has decreased by a significant amount, where ``significant'' is decided by the rule,
\begin{equation}
\Sigma_{i=1}^{N_1}|f_i| -\Sigma_{i=1}^{N_2} |f_i| \ge 
N_{c} \sqrt{\Sigma_{i=1}^{N_1} {\sigma_{f_i}}^2 },
\label{eq-ca}
\end{equation}
where $f_i$ is the magnetic flux of the $i$-th feature, 
$\sigma_{f_i}$ is the noise level in the flux of the $i$-th feature,
$N_1$ is the number of features involved in a cancellation event
at time step $t_1$, $N_2$ is the number of features involved in the
same cancellation event at time step $t_2$ and 
$N_{c}$ is an arbitrary factor (here we chose $N_{c}$=5).

We have three types of cancellation events.\\

\noindent
(a) Complete cancellation: If at any time step $t_1$, a spatial overlap (after dilation, 
see Appendix~\ref{appendixc}) of two sets 
of opposite polarity features leads to disappearance of both sets of features, 
then it is marked as a complete cancellation event.\\

\noindent
(b) Semi cancellation : If at any time step $t_1$, a spatial overlap (after dilation, 
see Appendix~\ref{appendixc}) of two sets of opposite 
polarity features is found to satisfy the condition for cancellation
(see above) according to Equation~(\ref{eq-ca}),
leading to disappearance of features of only one polarity, then the event 
is considered to be a semi cancellation event. \\

\noindent
(c) Partial cancellation : If at time step $t_1$, a spatial overlap (after dilation, 
see Appendix~\ref{appendixc}) of two sets of opposite
polarity features is found to satisfy the condition for cancellation
(see above) as described in Equation~(\ref{eq-ca}), leading
to survival of both the polarity features, then the
event is identified as a partial cancellation event.\\

\noindent
$\bullet$ Instantaneous (magnetic) flux of a feature: The flux of a feature at 
a given time step. The instantaneous flux of a feature in a birth event 
(death event) is the flux that it carries at the time step of its birth (death).\\

\noindent
$\bullet$ Maximum (magnetic) flux of a feature: The maximum value of flux of a feature 
attained during the course of the feature's lifetime.\\

\noindent
$\bullet$ Total instantaneous (magnetic) flux in a birth or a death event: The sum of 
the fluxes of all those features in the entire considered data set, which are 
born by a particular birth type or died by a particular death type at the 
time of birth or time of death (sum of instantaneous fluxes). \\ 

\noindent
$\bullet$ Total maximum (magnetic) flux in a birth or a death event: The sum of 
the maximum fluxes of all those features in the entire data set, 
which are born by a particular birth type or died by a particular death type.\\ 

\noindent
$\bullet$ Total (magnetic) flux (in emergence or cancellation): The sum of
the (instantaneous or maximum) fluxes of those features in the time series,
that participate in emergence or cancellation. Here the total flux in emergence 
includes not just the flux of features that are born by emergence, but the total flux emerged, 
which is $\Sigma F_{\rm {SYM,EM}}+F_{\rm {ASYM,EM}}$, with $F_{\rm {SYM,EM}}$ and $F_{\rm {ASYM,EM}}$ 
defined above. Similarly the total flux
in cancellation not only includes the flux lost in features that die by cancellation, but 
also includes the flux lost in surviving features in semi and partial cancellation.\\

\noindent
$\bullet$ Total (magnetic) flux (in all the birth events): The sum of 
the (instantaneous or maximum) fluxes of those features in the time series, 
that are born in all considered birth events. In case of total instantaneous flux, 
it is the sum of the fluxes at the time of birth.\\

\noindent
$\bullet$ Total (magnetic) flux (in all the death events): The sum of 
the (instantaneous or maximum) fluxes of those features in the time series, 
that died from all considered death events. In the case of the total instantaneous 
flux, it is the sum of fluxes at the time of death. \\


\begin{figure*}[ht]
\sidecaption
\includegraphics[width=6cm]{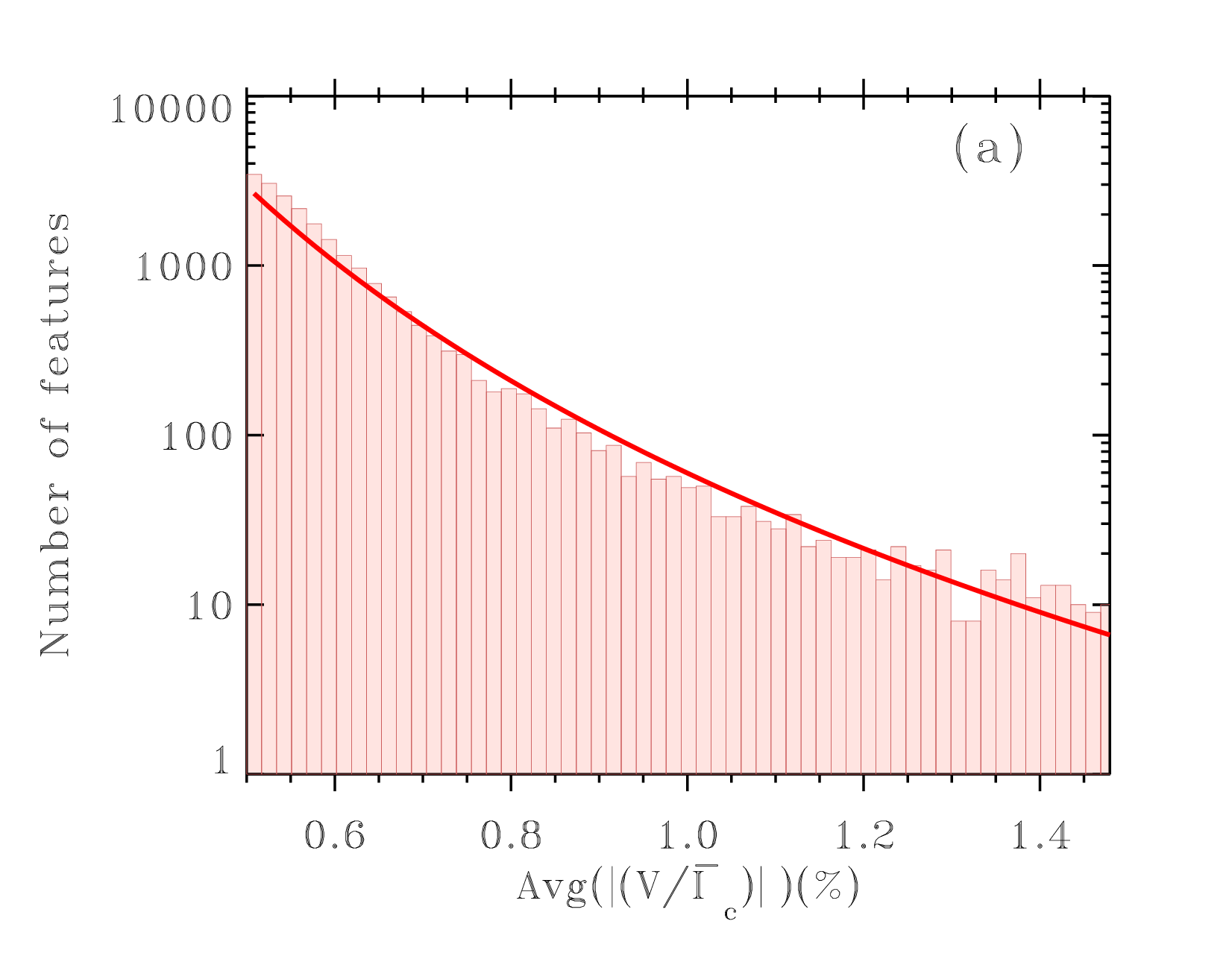}
\includegraphics[width=6cm]{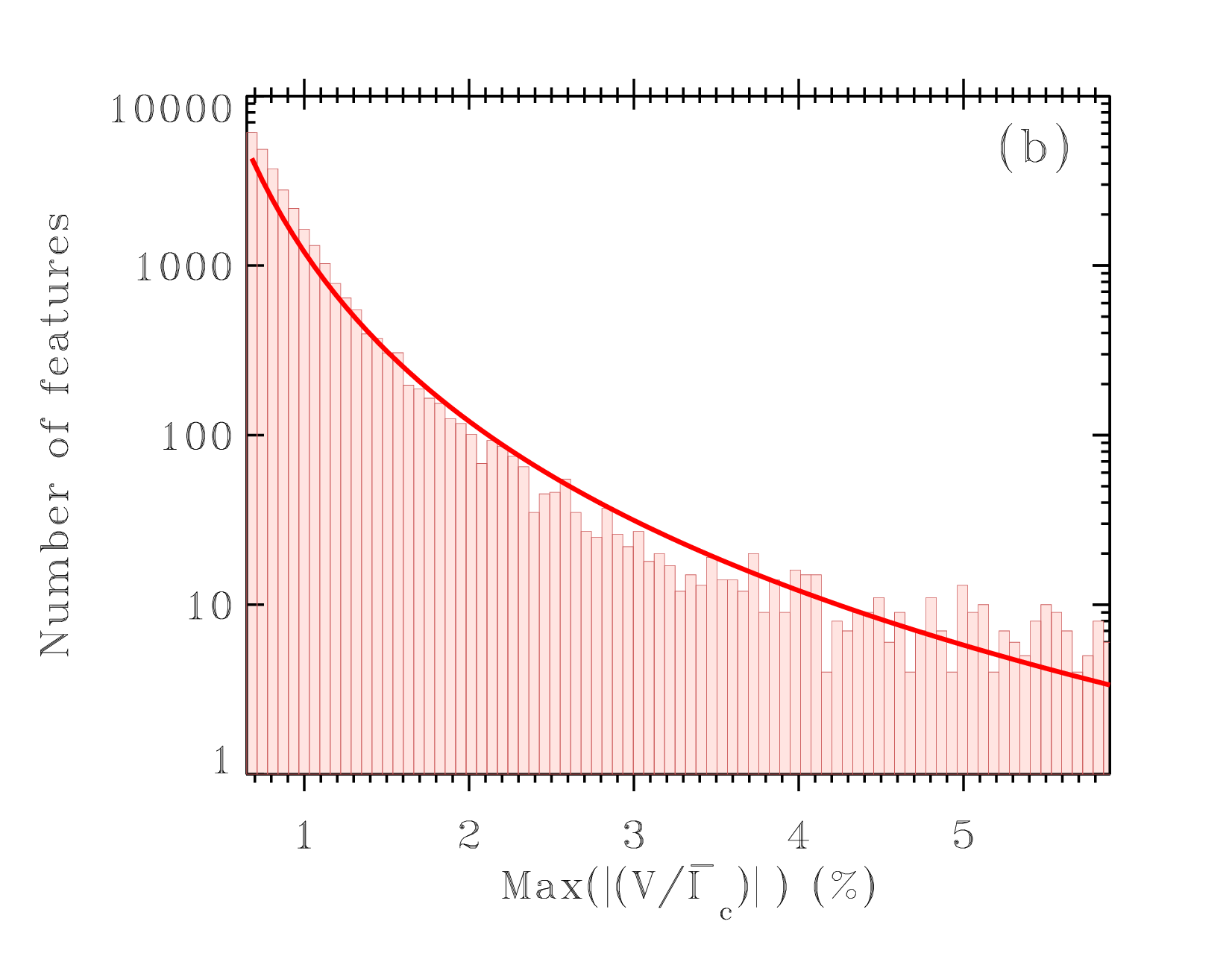}
\caption{Histograms of magnetic-feature averaged value of $|V/\bar{I}_c|$ (panel a) and
maximum value of $|V/\bar{I}_c|$ per feature (panel b). Red lines are fitted power-law 
curves to these histograms (see text).}
\label{fig6}
\end{figure*}

\section{Results and discussion}
\label{results}

\subsection{The distributions of average and maximum value of $|V/\bar{I}_c|$}
In Fig.~\ref{fig6} we plot histograms of average 
and maximum of the $|V/\bar{I}_c|$ within each feature.
Note that smaller values are probably affected by the 
noise, and the finite spatial resolution. The average $|V/\bar{I}_c|$ distribution has 
mean and median values of $0.51$\,\% and $0.48$\,\% respectively, and
the maximum $|V/\bar{I}_c|$ has mean and median values of $0.82$\,\% and $0.69$\,\% 
respectively. The average and maximum values of $|V/\bar{I}_c|$
follow power-law distributions with power law indices $-4.6$ and $-3.3$ 
respectively. These indices are much larger than those found for the 
magnetic flux per feature \citep[see Sect.~\ref{area-flux-blos};][]{parnelletal09,buehleretal2013}.

\begin{figure*}[ht]
\sidecaption
\includegraphics[width=6cm]{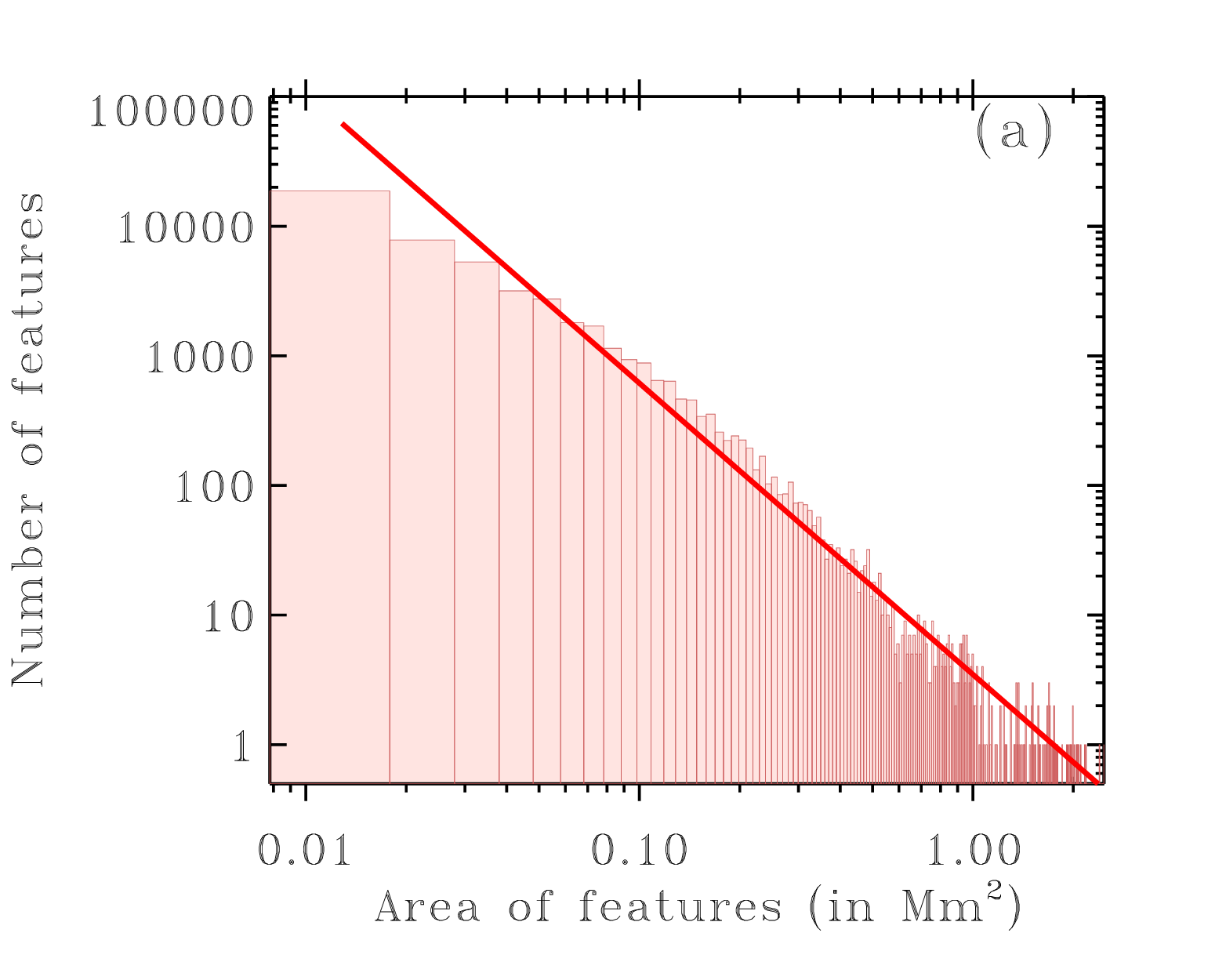}
\includegraphics[width=6cm]{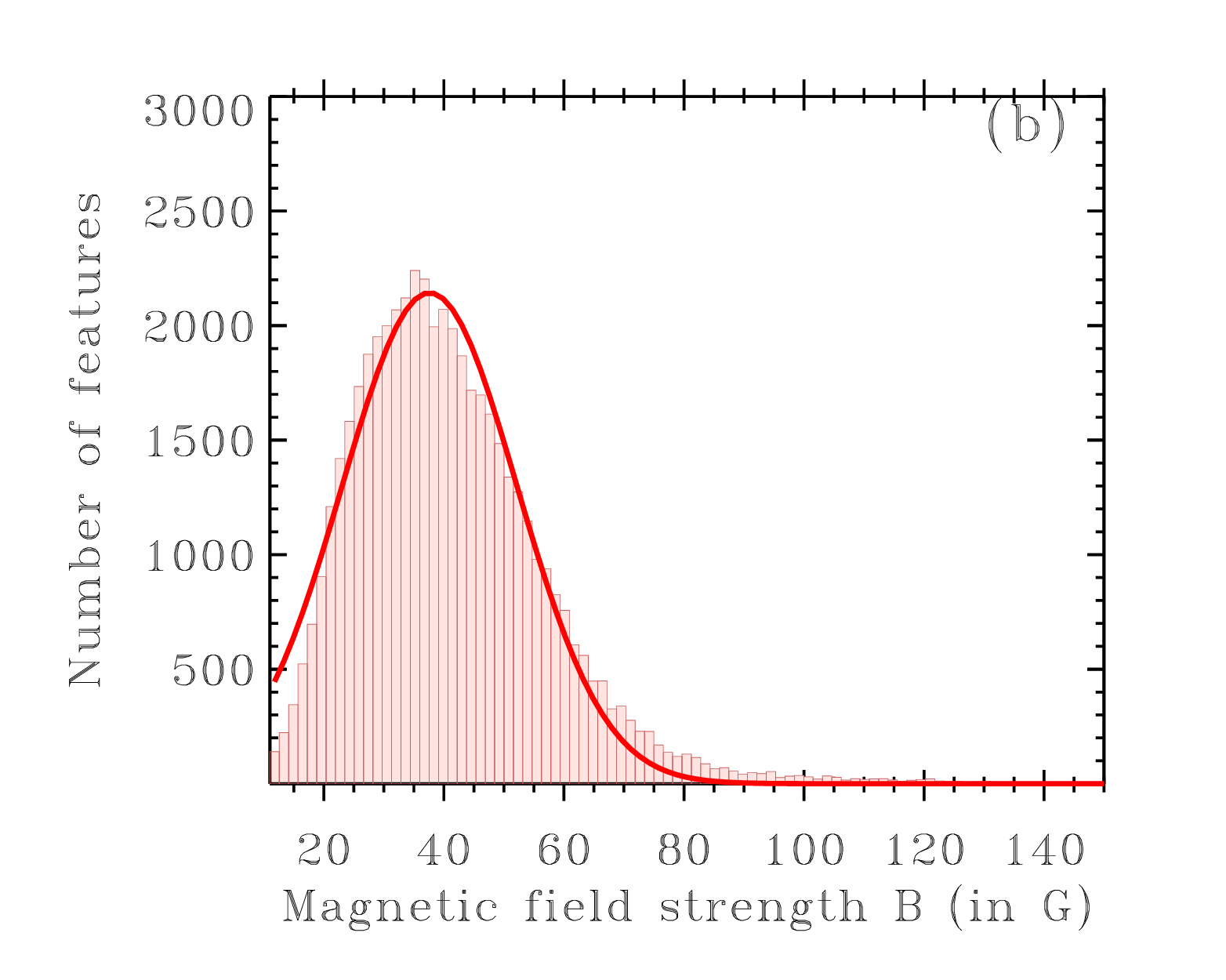}
\caption{Histograms of areas (panel a) and feature-averaged LOS magnetic field values 
(panel b) in the features. Red lines indicate a power-law fit in panel (a) and a 
normal distribution in panel (b).}
\label{fig5}
\end{figure*}

\begin{figure*}[ht]
\sidecaption
\includegraphics[width=6cm]{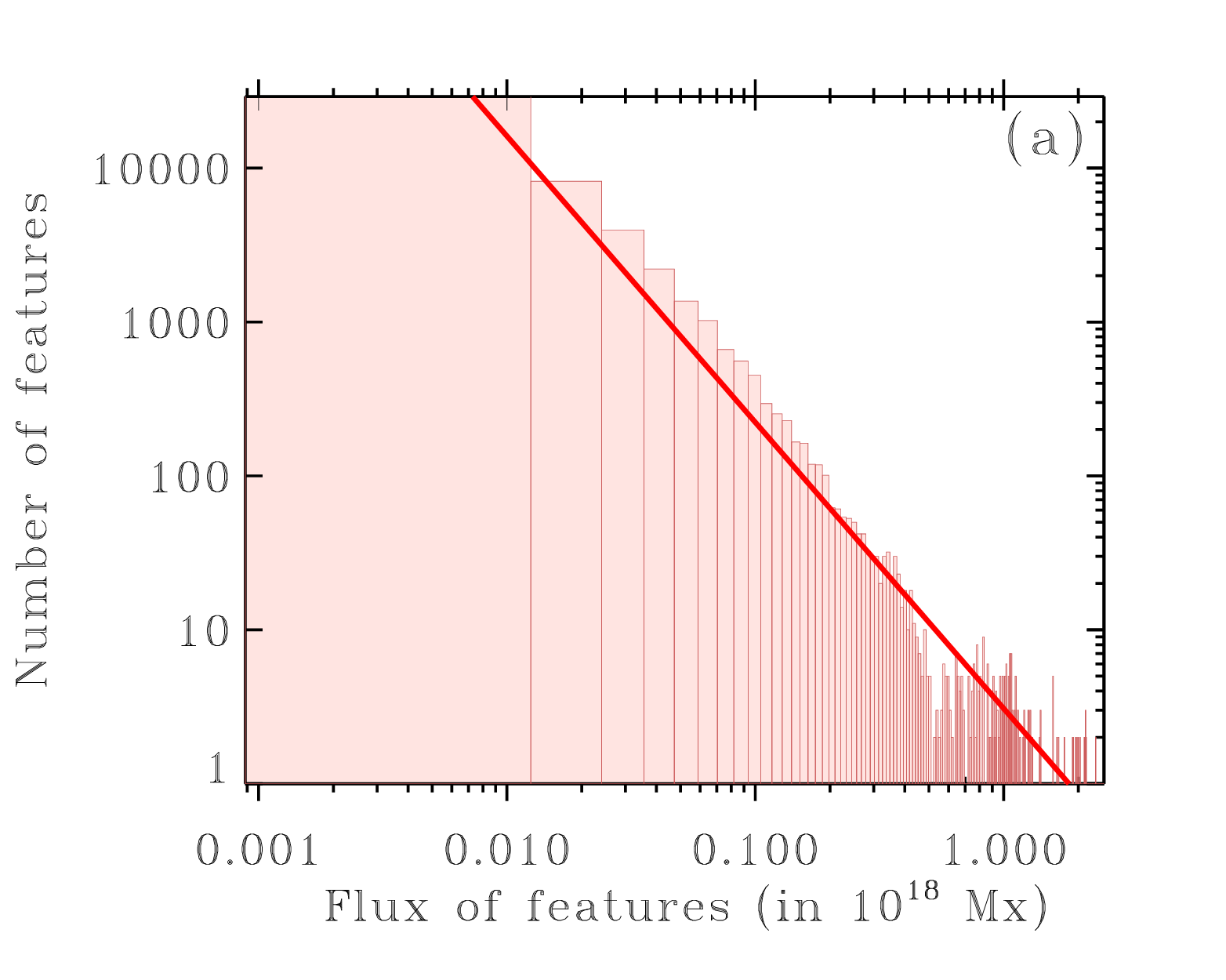}
\includegraphics[width=6cm]{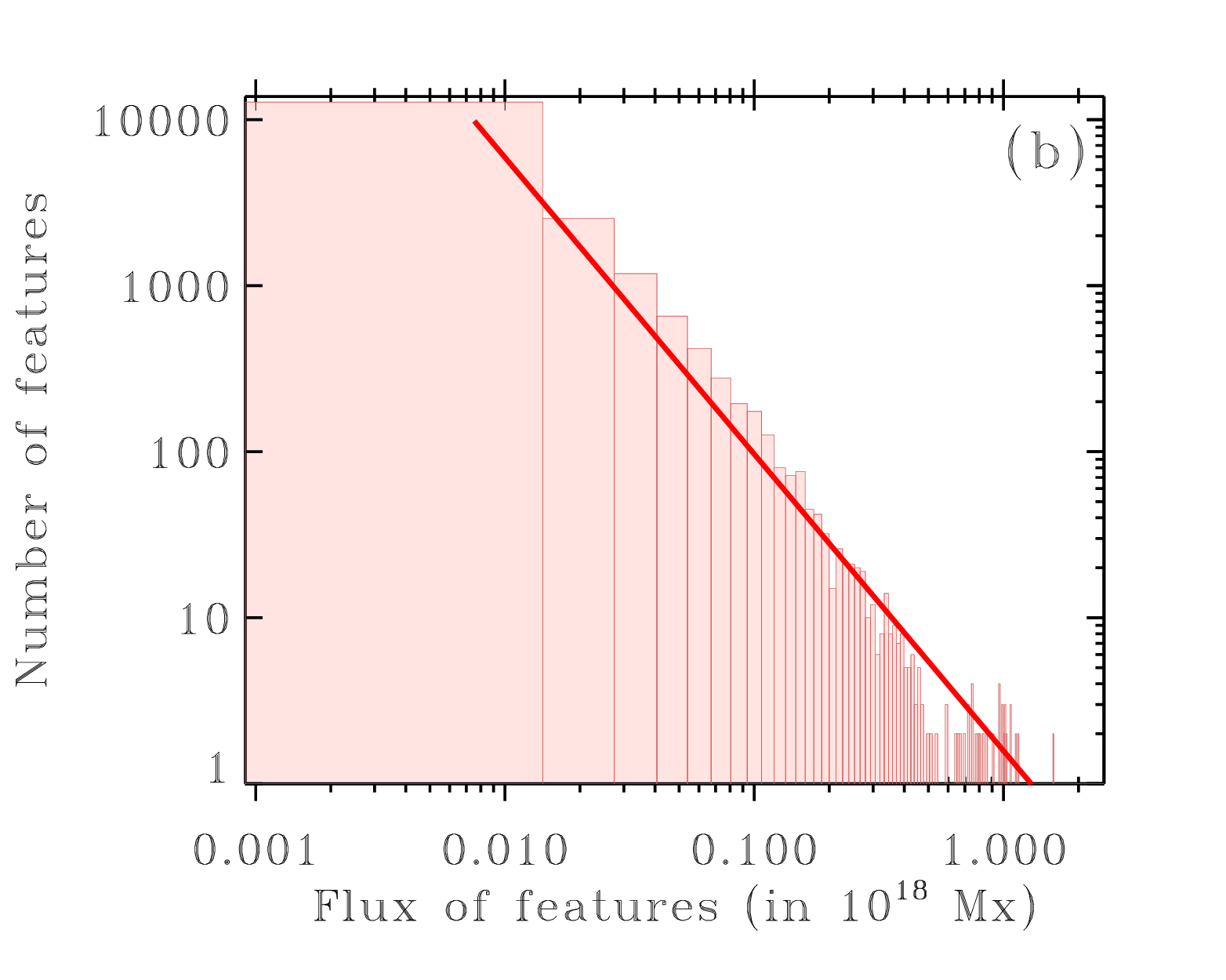}
\caption{Histograms of instantaneous (panel a) and maximum (panel b) 
fluxes of the features. Red lines indicate power-law fits.}
\label{fig5a}
\end{figure*}

\begin{figure*}[ht]
\sidecaption
\includegraphics[width=6cm]{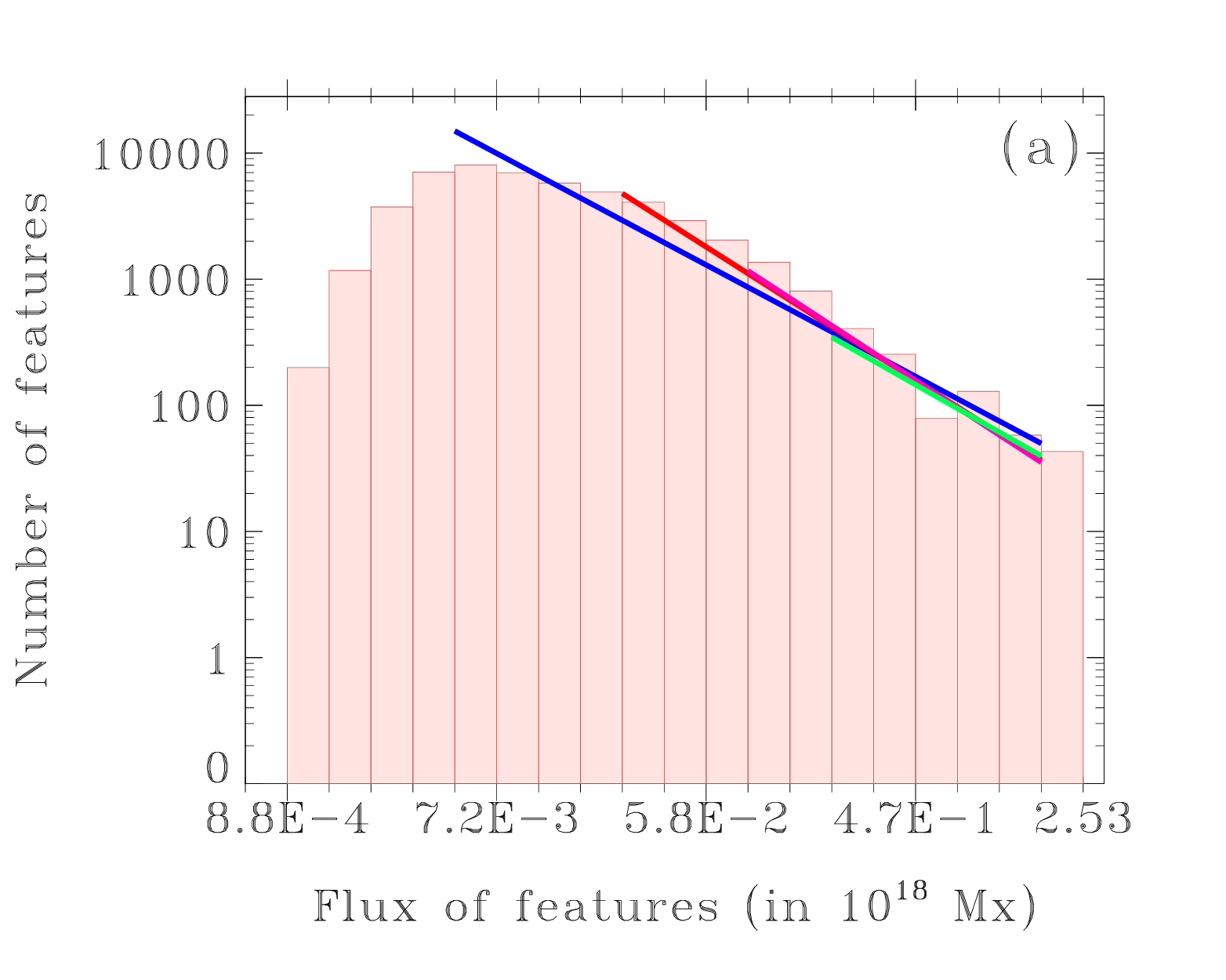}
\includegraphics[width=6cm]{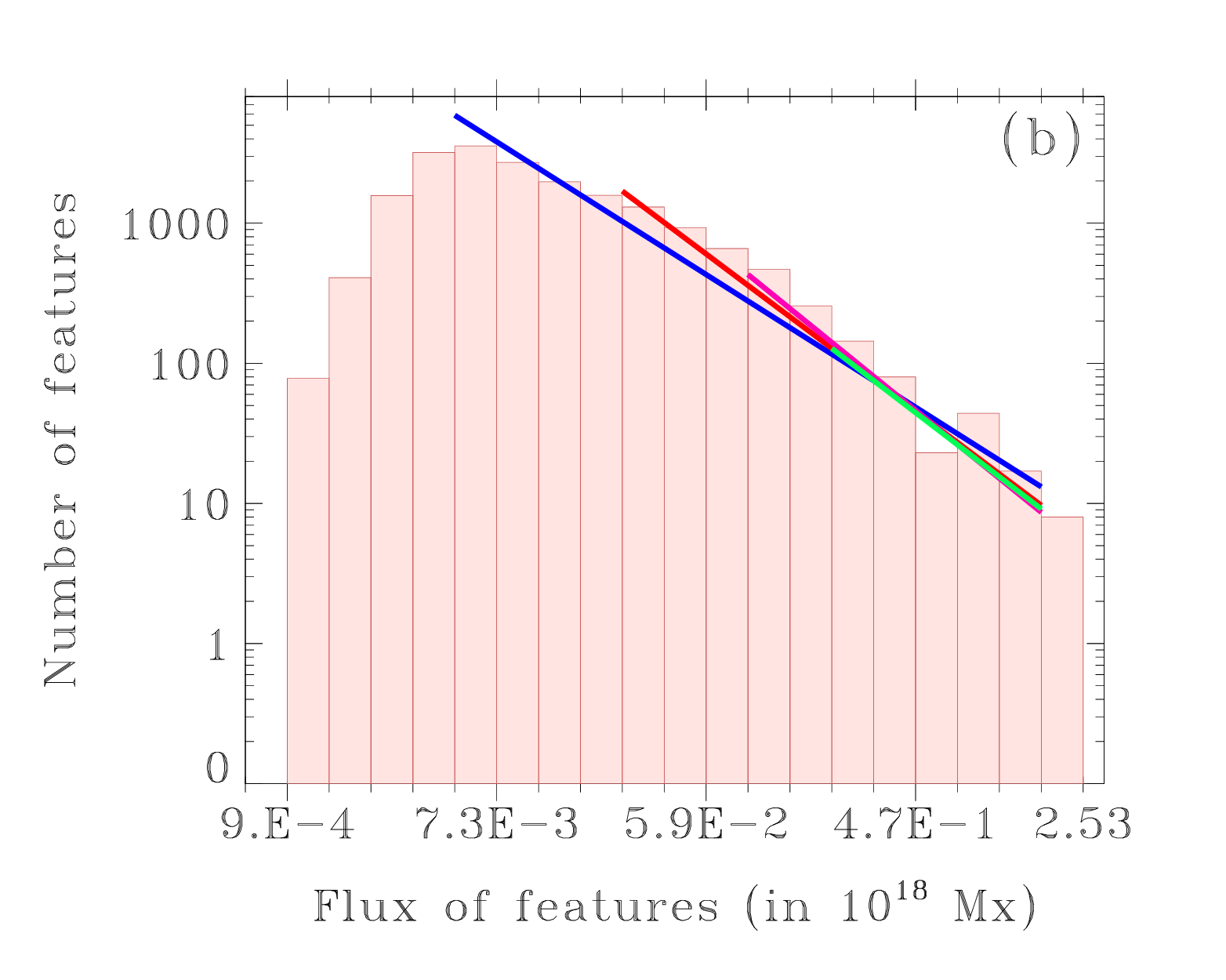}
\caption{Histograms of instantaneous (panel a) and maximum (panel b) 
of the fluxes of the features, with bins equi-distant in log(flux). 
Colored lines indicate power-law fits with different minimum 
fluxes given in table~\ref{table_1}.}
\label{fig5b}
\end{figure*}

\subsection{Distribution of the area, magnetic field and magnetic flux of the features}
\label{area-flux-blos}
In Fig.~\ref{fig5}(a) we plot the histogram of magnetic feature area. 
The areas of the features range from a minimum of
$\sim 7.8 \times 10^{-3}$ Mm$^2$ (or 5 pixels, which is the lower limit we 
have set for features selected for investigation) to a maximum of 
$\sim$ 2.5 Mm$^2$ ($1585$ pixels). The areas of the magnetic features closely 
follow a power law distribution of the form $a\,x^b$ with index $b=-2.25$. It closely matches with 
the power law index of $-2.2$ obtained by other recent studies using 
Hinode/SP data for such area distributions \citep[see][and the references 
cited therein]{buehleretal2013}.

A histogram of the feature-averaged LOS magnetic field component, determined 
using the COG technique (see section~\ref{flux-computation} for 
details), is displayed in Fig.~\ref{fig5}(b). 
The magnetic field values thus obtained are weak, as expected from 
Fig.~\ref{fig-flux-calib2}, and the distribution is surprisingly symmetric,
although with a tail towards stronger fields. The 
mean and median values are $41$\,G and $39$\,G respectively.
The feature-averaged magnetic field strength values
range between $11$\,G to $198$\,G. These values are somewhat lower than expected for 
an equipartition field \citep[200-400 G, e.g.,][]{solankietal96}. 
The field strength values nearly follow a normal distribution.

In Fig.~\ref{fig5a}(a) we plot the histograms of the instantaneous fluxes 
of the magnetic features. Here we consider all the 50255 features 
in the time series, i.e. considering all the features that represent evolution/history 
of a given feature in different time steps, to be independent.
The minimum and maximum of the distribution is $9\times10^{14}$ 
Mx and $2.5\times10^{18}$ Mx respectively. 
The fluxes follow a power law distribution of the 
form $a\,x^b$ with index $b=-1.85$ 
which is consistent with 
\citet[][]{parnelletal09}.
Other recent studies using Hinode data also obtain a power law index
close to $-1.8$ \citep[see e.g.,][the focus of the former being 
the network fields]{iidaetal2012,buehleretal2013}.

In Fig.~\ref{fig5a}(b) we plot histograms of maximum fluxes                    
of the magnetic features. Here maximum flux of a feature refers to the maximum of 
the feature's magnetic flux during it's entire lifetime (see Sect.~\ref{def}), with 
lifetime being obtained for the 10:1 area-ratio criterion. 
The maximum fluxes also follow a power-law distribution with power-law index $b=-1.78$.
The power-law index for maximum fluxes is close to that obtained for instantaneous
fluxes as are the minimum and maximum flux values. The fit parameters are listed 
in Table~\ref{table_1}.

However, the tail region of the histogram suffers from very poor statistics in
Fig.~\ref{fig5a} leading to seeming gaps in the flux per feature. 
To avoid these holes, we plotted the histograms with bins 
of equal width in logarithm of the fluxes in Fig.~\ref{fig5b}. Power-law fits to 
the histograms of the fluxes depend sensitively on the smallest flux per feature in the  
distribution \citep[see also][]{parnelletal09,iidaetal2012}. 
We show power-law fits of these histograms for four different smallest flux per feature values.
The fit parameters of these distributions, and the corresponding $\chi^2$ 
values are listed in Table~\ref{table_1}. We note here that the value $2\times10^{17}$ Mx 
for the smallest flux per feature, chosen for the green power law fit in Fig.~\ref{fig5a} 
is same as that in \citet[][]{parnelletal09} for the Hinode data. 
In the remaining cases, the two extreme smallest flux per feature, 
either try to fit numbers of features that are clearly 
affected by instrumental resolution and sensitivity, or affected by the small number of 
points it is fitting. However, they give an indication of the
uncertainty of the power-law index to the choice of smallest flux per feature. 
Therefore, the fit with smallest flux per feature of $10^{16}$ Mx
appears to provide the most reasonable fit among the four cases. 
Clearly, the power-law index depends strongly on the type of binning, with 
all the fits to the bins of equal size on a logarithmic scale giving far 
lower power-law indices than those to the bins of equal size in linear scale. 
To our knowledge this dependence has not been previously highlighted 
and the type of binning is often not discussed carefully in studies of magnetic flux distribution. 
There is a need to investigate these effects thoroughly, as having bins that are equally 
broad on a logarithmic scale returns power-law fits that are not consistent with the results of 
\citet[][]{parnelletal09}. At present it is unclear, which type of binning provides the better description. 
However, a study of the best type of binning is outside the scope of this paper.

The feature-averaged field strength values shown in this paper are much 
smaller compared to the field strengths of quiet Sun features 
found by \citet[][]{khomenkoetal2003,laggetal2010}
and \citet[][]{riethmulleretal2014}.
The main reason for this discrepancy is that we show feature-averaged values. Also, we 
determined only the LOS component of the field and, unlike the latter two groups of authors 
cited above, consider all magnetic features visible in Stokes $V$ and not just those 
that are associated with bright points. In Fig.~\ref{fig-flux-calib1} we 
have shown the flux of individual pixels where 
the $B_{\textrm{LOS}}$ reach kG values.

\begin{figure}
\centering
\includegraphics[scale=0.45]{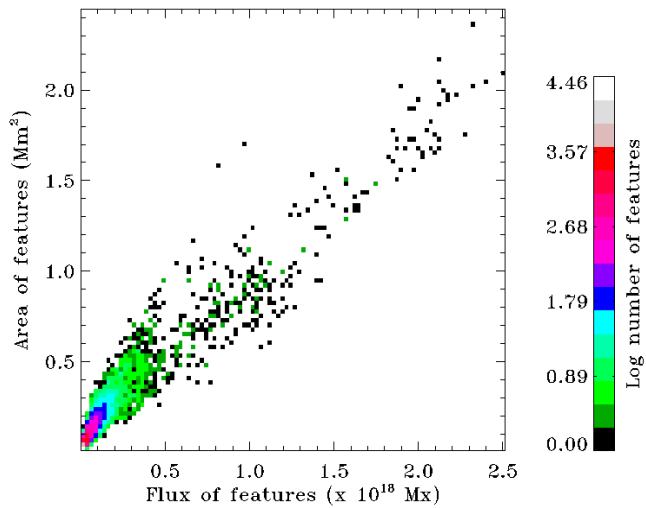}
\caption{2D histogram of the number of
features on the plane spanned by the magnetic flux $\Phi$ in the features
and their areas $A$.}
\label{fig-scat2}
\end{figure}

In Fig.~\ref{fig-scat2} we present a 2D histogram of the number of
features on the plane spanned by the magnetic flux $\Phi$ in the features 
and their areas $A$. This figure shows that area and fluxes are almost linearly
related. This linear behavior is useful because it implies that we can constrain 
the classification of birth and death events, either through an area-ratio criterion or 
through a flux ratio criterion according to convenience.

\subsection{Birth and death of magnetic features}
\label{birthanddeath}
In this section we discuss statistics of the births and deaths of magnetic features. We
start by considering only interactions between features of the same polarity in Sect.~\ref{unipol}.
Later, in Sect.~\ref{bipol} we include also interactions between magnetic features of 
opposite polarity.
\subsubsection{Unipolar interactions}
\label{unipol}
In Table~\ref{table_2} we present the number and percentage of features that are born or die 
according to different processes. The numbers are given for four different 
area-ratio criteria (see Sect.~\ref{def} for details).
For the 10:1 area-ratio criterion, a total of 18037 features were 
born{\footnote{Here 18037 features corresponds to the total number of features born, excluding the 
number of features alive in first time frame.}} 
in the course of 40 time steps (i.e. over the entire considered time series),
while $17779$ features died. The difference between the births and deaths is explained by
the fact that whereas $1019$ features were present in the first frame, $1277$ were there in the last.
As we decrease the area-ratio criterion, from 10:1 to 
2:1, i.e. tighten the requirement for a merging or splitting event to result in a birth or death, 
the number of features that are born is reduced to $15124$ and those that die to $14866$.
We have splitting-off birth and merging-into death processes resulting from those features
which do not satisfy the area-ratio criterion in splitting and merging (see Sect.~\ref{def} for
details). They form a significant fraction of features that are born or have died. As expected, 
this number becomes largest for the extreme area-ratio criterion of 2:1.

Among the other types of unipolar birth events, most features are born due to 
appearance, followed by splitting, and then by merging. Not surprisingly, the number 
of features born due to splitting is more than twice that from merging, since splitting 
gives rise to the birth of two or more features whereas merging leads to the birth of 
only one feature. With a decrease in the area-ratio criterion, the number of features born by 
splitting and merging decreases, those born by splitting-off birth events increases, 
since the features that would otherwise result from splitting and merging are now 
considered to be simply born by splitting-off birth events.
Similarly, among the unipolar death events, 
death due to disappearance is the most common among processes responsible for the 
death of magnetic features, followed by merging and then by splitting. 
For the same reason that there are nearly twice as many births by splitting as by 
merging, there are nearly twice as many deaths by merging than by splitting. 

In Table~\ref{table_3} we list the total instantaneous magnetic flux of the features 
in various birth and death events (see Sect.~\ref{def}). About $3.6 \times 10^{18}$ Mx of flux is present at the
beginning of the time-series and nearly the same amount of flux ($4.0 \times 10^{18}$ Mx)
exists at the end of the time-series. 
The amount of flux is distributed differently between the various birth processes
than just the number of events. E.g., for the area-ratio criterion of 10:1, 
although nearly half of the features are born by appearance, this process
contributes only $12$\,\% to the total flux in newborn features. This suggests that mainly small
features are born by appearance which is confirmed by considering the flux per 
feature also listed in Table~\ref{table_3}. Splitting and merging events, respectively, provide 
$44.5$\,\% and $41.5$\,\% of the total flux of new features (although not new flux at
the solar surface, since fresh flux is provided only by appearance and 
emergence events). The percentage of the flux in appearance events increases as the 
area-ratio criterion is tightened (i.e. the ratio made smaller) and reaches 
$27$\,\% of the total instantaneous flux for the 2:1 area-ratio case. The flux contained in 
splitting-off birth events are as small as $1.5$\,\% for an area-ratio of 10:1, which increases
with a decrease in the area-ratio and reaches $14.6$\,\% for the 2:1 case.
A similar behavior is observed in death events.

In Table~\ref{table_4} we provide the total maximum flux 
distribution in various birth and death events (see Sect.~\ref{def}). 
The total maximum flux has a slightly different distribution than the 
total instantaneous flux in different birth and death events. As discussed 
above, the total instantaneous flux distribution in birth events shows that features 
born by appearance have lower values of flux on average. Whereas, 
the maximum flux in a feature born by appearance reached in the course of its life is 
roughly a factor of nearly $1.5-1.6$ larger than its flux at birth 
\citep[see also][]{gosicetal2016}.
Consequently, features born by appearance events display a larger relative
contribution to the total flux when the maximum flux is counted.
Conversely, the contribution of features born by splitting and merging 
to the total maximum flux is smaller than their contribution to the 
total instantaneous flux.
A similar behavior is observed in death events.

\subsubsection{Interactions between features with opposite polarity}
\label{bipol}
In this section we investigate the contribution to the total flux from emergence
and cancellation of bi-polar magnetic features. In the case of emergence, we consider different 
flux-ratio criteria such that only those events are considered to be bi-polar emergences 
in which the ratio of fluxes between the involved opposite polarity features is smaller 
than the imposed ratio. Here we consider the ratios 10:1, 5:1, 3:1 and 2:1. However, since
flux and area are linearly related (see Fig.~\ref{fig-scat2}), hereafter,
to avoid confusion in discussions, we use the term ``area-ratio criterion'' in place of 
flux-ratio criterion. We do not consider any flux-ratio criterion for cancellation events, 
i.e. the flux in the two sets of the two opposite polarity features involved in the 
cancellation can be arbitrarily different, thus allowing for partial cancellation 
(see definition of cancellation events in 
Sect.~\ref{def}).\footnote{The total flux in the death events is the sum of fluxes of the 
features that died from all the unipolar and bi-polar 
death events, which depends on the area-ratio criterion. Therefore, when the fluxes 
are normalized by the total flux, the fractional fluxes depend on the
area-ratio criterion.}

Table~\ref{table_2} reveals that only $1-2$\,\% of the features in the 
considered data set are born through bi-polar emergence, and at the time of birth 
these features carry only $0.3-0.5$\,\% of the total instantaneous magnetic flux 
in the birth events. When we consider total maximum magnetic flux in birth events, 
they contribute between $0.5-1.0$\,\%. Thus emerging features grow to reach $1.7-2$ times 
their flux at birth. We note here that, the fluxes given here and
in Tables \ref{table_3} and \ref{table_4} correspond to the fluxes of the 
features that are born due to emergence. I.e. these include the fluxes of the 
features with both polarities in time-symmetric emergence events, as well as the 
fluxes of the newly emerged features in time-asymmetric emergence events (but 
not the fluxes of the already existing opposite-polarity features, since these are 
not newly born features). The total emerged flux in all the emergence events, namely, 
$\Sigma (F_{\rm {SYM,EM}}+F_{\rm {ASYM,EM}})$ (as defined in Sect.~\ref{def}) is higher than 
these values, in both, instantaneous and maximum flux. 
Since the asymmetric emergence dominates over symmetric emergence, the total flux that 
emerged is nearly twice the total flux of the features that are born due to emergence. 
For the 10:1 area-ratio criterion, a total of $3.9\times10^{18}$ Mx instantaneous
flux and $9\times10^{18}$ Mx maximum flux is gained in emergence.
Even then the total flux in emergence events is an order of magnitude smaller compared
to the total flux detected in appearances. Hence either bi-polar emergence events
are difficult to detect in the quiet Sun, or these events are comparatively rare. 

On the other hand we see nearly $8.7-10.4$\,\% of the features dying due to cancellation. 
The total instantaneous flux lost in cancellation events ranges between $1.7$\,\% and $3.6$\,\% 
of the total flux lost in death events. When we consider total maximum flux, it 
ranges between $2.9$\,\% and $5.6$\,\%. We note that the fluxes ascribed to cancelling 
events given here and in Tables \ref{table_3} and \ref{table_4} correspond only to the fluxes of the 
features which undergo cancellation and die.
I.e. the fluxes lost from the interacting features in semi and partial cancellation 
events which survive in the next time step, are not included here.
This is because here we are interested in the fluxes that account for death of the 
features. The total flux lost in complete, semi and partial cancellation 
events (i.e. the sum of fluxes lost in died and surviving features) is nearly four times the 
total flux in the features that die due to complete, semi and partial cancellation.
This accounts for $7$\,\% of the total instantaneous flux and $11$\,\% of the 
total maximum flux loss in cancellation for the 10:1 area-ratio criterion. 
In this case, a total of $3.1\times10^{19}$ Mx 
instantaneous flux and $5.3\times10^{19}$ Mx maximum flux is lost. A smaller number of 
cancelling features contributing more flux loss suggests that the features undergoing 
cancellation are on average large. However, these large cancelling features are the features
that survive after semi and partial cancellation because cancelling features that disappear are
small and carry low flux (see discussions in Sect.~\ref{evolution}).
The total instantaneous flux (total maximum flux) that is lost in the 
cancellation is nearly $8$ times ($6$ times) larger than the total flux gained in emergence for 
the 10:1 area-ratio criterion. Also, a roughly equal amount of flux is removed 
from the magnetograms (and possibly from the solar surface) by cancellation as by 
disappearance.

In Table~\ref{table_5} we list the flux distributions in time-symmetric
and time-asymmetric emergence events with respect to the total flux in 
emergence events (not with respect to the total flux in all the birth events), 
as well as in complete, semi and partial cancellations with respect to only the total flux
lost in cancellation.
For example, in the case of 10:1 area-ratio criterion, the fractions of the total emerged 
instantaneous flux of $3.9\times10^{18}$ Mx in time-symmetric and time-asymmetric
emergence events is listed in the 3rd and 4th rows of Table~\ref{table_5}. 
It is interesting to note that, among the captured bi-polar interactions,
$80-90$\,\% of the features emerged in time-asymmetric emergence events. 
The time-symmetric emergence events corresponds to 
only $3-10$\,\%, ($4-14$\,\%) of the total instantaneous (total maximum) flux
added by bi-polar emergence, so that time-asymmetric emergence events dominate.
In cancellation events, complete cancellation corresponds to only $2$\,\% ($3$\,\%) of 
the total instantaneous (total maximum) flux that is lost in cancellations 
while semi and partial cancellation events account for the rest.

\subsection{Discussion}
\label{disc}
A similar study as described in Sect.~\ref{birthanddeath} has been carried out
by \citet[][]{lambetal2008} based on SOHO MDI magnetograph data, and by
\citet[][]{zhouetal2010,wangetal2012,lambetal2013,gosicetal2014,gosicetal2016} 
based on Hinode/SOT/NFI data. 
Their approach differs somewhat from ours in the definition of the events and the 
type of data used and partly also in the aims of the studies. 
Thus, our data have higher spatial resolution, so that
we see smaller features. Also, these authors did not
introduce a threshold for merging and splitting events. A comparison can still be 
useful, however, although it only makes sense to compare the statistics obtained with 
10:1 area-ratio criterion. In the following we provide a detailed comparison of 
our results with those by \citet[][]{lambetal2008,lambetal2013}.\\

The following discussion is summarized in Table~\ref{table_6}.\\

\noindent
(1) The appearances account for $12$\,\% of the total flux found by 
\citet[][]{lambetal2008}. The flux contained in appearance in the present work
accounts for $12$\,\% of the 
total instantaneous flux and $19.1$\,\% of the total maximum 
flux for area-ratio criterion 10:1. Since their fluxes correspond to the 
maximum flux, this shows that we obtain more total flux in 
appearance events than they find.\\

\noindent
(2) The `fragmentation' birth type in \citet[][]{lambetal2008}
gives rise to $76$\,\% of the total flux in newborn features, whereas we find 
$44.5$\,\% and $41.5$\,\% of total instantaneous flux and $41.8$\,\% and 
$36.5$\,\% of total maximum flux in splitting and merging birth events, 
respectively, giving a total of $80$\,\%. The small difference 
between the two results is partly due to differences in the details of the definitions. \\

\noindent
(3) Emergence accounts for nearly $1$\,\% of the total flux in \citet[][]{lambetal2008}.
We have $0.5$\,\% of total instantaneous flux, and $1$\,\% of total maximum 
flux in the features that are born by emergence. However, as discussed earlier, the
actual total flux emerged is twice the total flux appeared in features born by emergence.
Therefore the total flux appearing through emergence accounts 
for $1-2$\,\% of the total flux in birth events, still in agreement with the 
result of \citet[][]{lambetal2008}. \\

\noindent
(4) When we focus only on the new flux at the solar surface, which means the total flux 
gained through appearance and emergence events, (i.e. excluding the total fluxes in 
the features born by splitting and merging events, which is not new at the solar surface), 
appearance accounts for $92.4$\,\% of the total instantaneous flux, and $91$\,\% 
of the total maximum flux, so that the new flux in bipolar emergence events accounts for less 
than $10$\,\% of the new flux at the solar surface, at least at the sensitivity and 
resolution of the Sunrise/IMaX instrument. In this point we assume that, the
term `total instantaneous/maximum flux' corresponds to the sum of the total instantaneous/maximum 
fluxes in all appearance and emergence events.\\

\noindent
(5) \citet[][]{lambetal2008} could not assign $10.9$\,\% of the birth events 
to a clear category (they referred to these as error events), while we uniquely
assigned a birth or death category to every feature as described in Section~\ref{def} and 
Appendix~\ref{appendixb}. This accounts for some of the differences in the results.\\

\noindent
(6) Uni-polar death types (disappearance, splitting and merging) are 
found to have similar statistics as birth types, in agreement with the findings of
\citet[][]{lambetal2008}. \\

\noindent
(7) In the analysis by \citet[][]{lambetal2013}, where the authors focus only
on disappearance and cancellation, nearly $83$\,\% of the total flux
is removed from the visible features by disappearance.
In our case, if we estimate the percentage of flux lost in disappearance and cancellation by 
considering only these two types of events, disappearances accounts for $59$\,\% 
of the total instantaneous flux, and $60$\,\% of the total maximum flux. 
However in the above estimation if we further leave out also the fluxes lost in 
the features that survive in semi and partial cancellation to consider only the 
fluxes lost in features that die by disappearance and cancellation 
\citep[as done by][]{lambetal2013} then disappearance accounts 
for $86$\,\% of the total instantaneous flux, and $85$\,\% of the total 
maximum flux, in good agreement with the results of \citet[][]{lambetal2013}. 
Here and in the next paragraph again, we assume that, the term `total instantaneous/maximum flux' 
corresponds to the sum of instantaneous/maximum fluxes in 
those events considered in the context.
\\

\noindent
(8) Cancellation in \citet[][]{lambetal2013} 
accounts for $12$\,\% of the total flux. If we leave out the 
fluxes contained in surviving features in semi and partial cancellation 
\citep[as done in][]{lambetal2013}, 
we obtain $14$\,\% total instantaneous flux and $15$\,\%
total maximum flux that corresponds to cancellation event, which is comparable
to the results obtained by \citet[][]{lambetal2013}. \\

\subsection{Lifetimes of the features}
\label{lifetime}
\begin{figure*}
\sidecaption
\includegraphics[width=6cm]{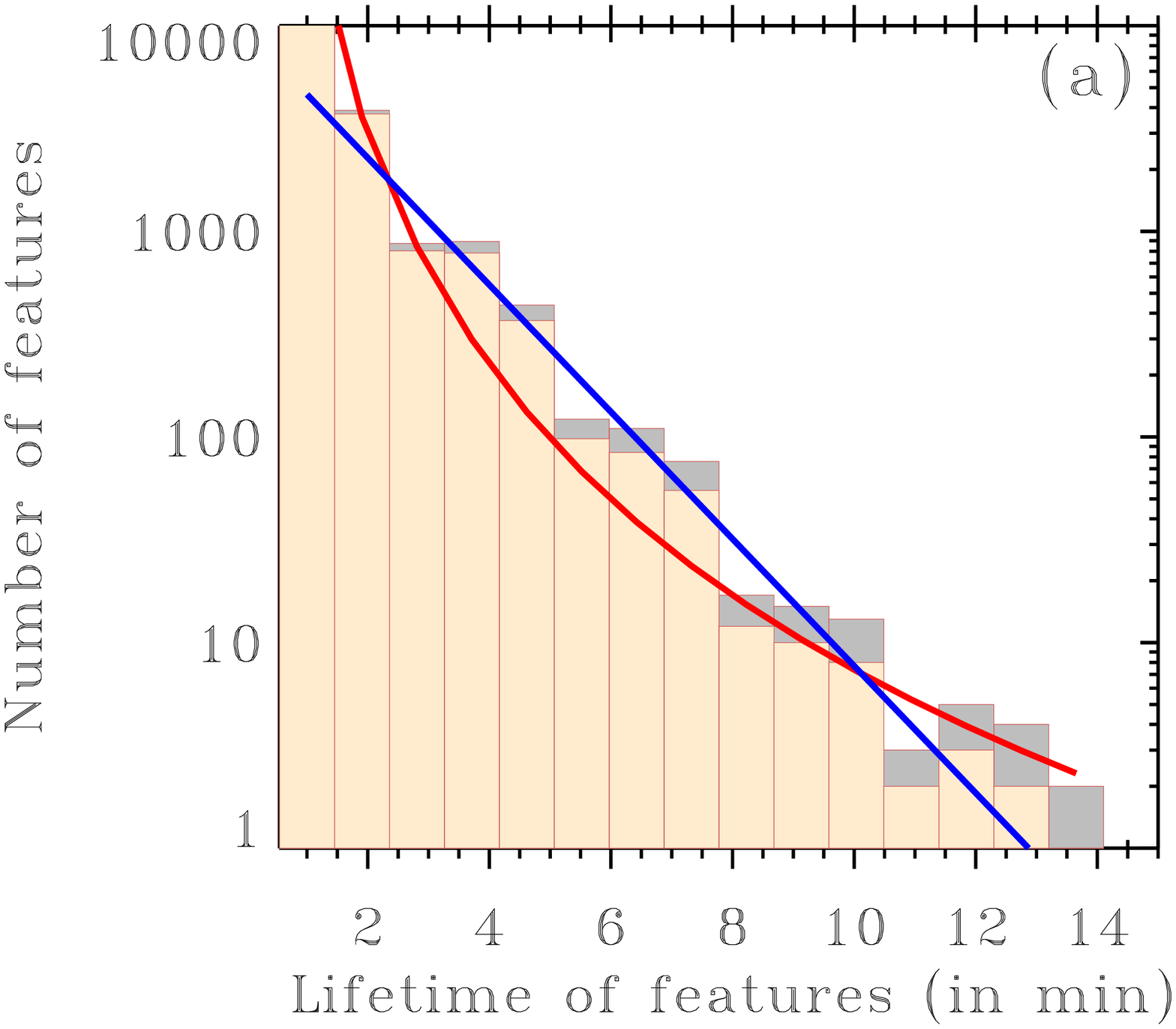}
\includegraphics[width=6cm]{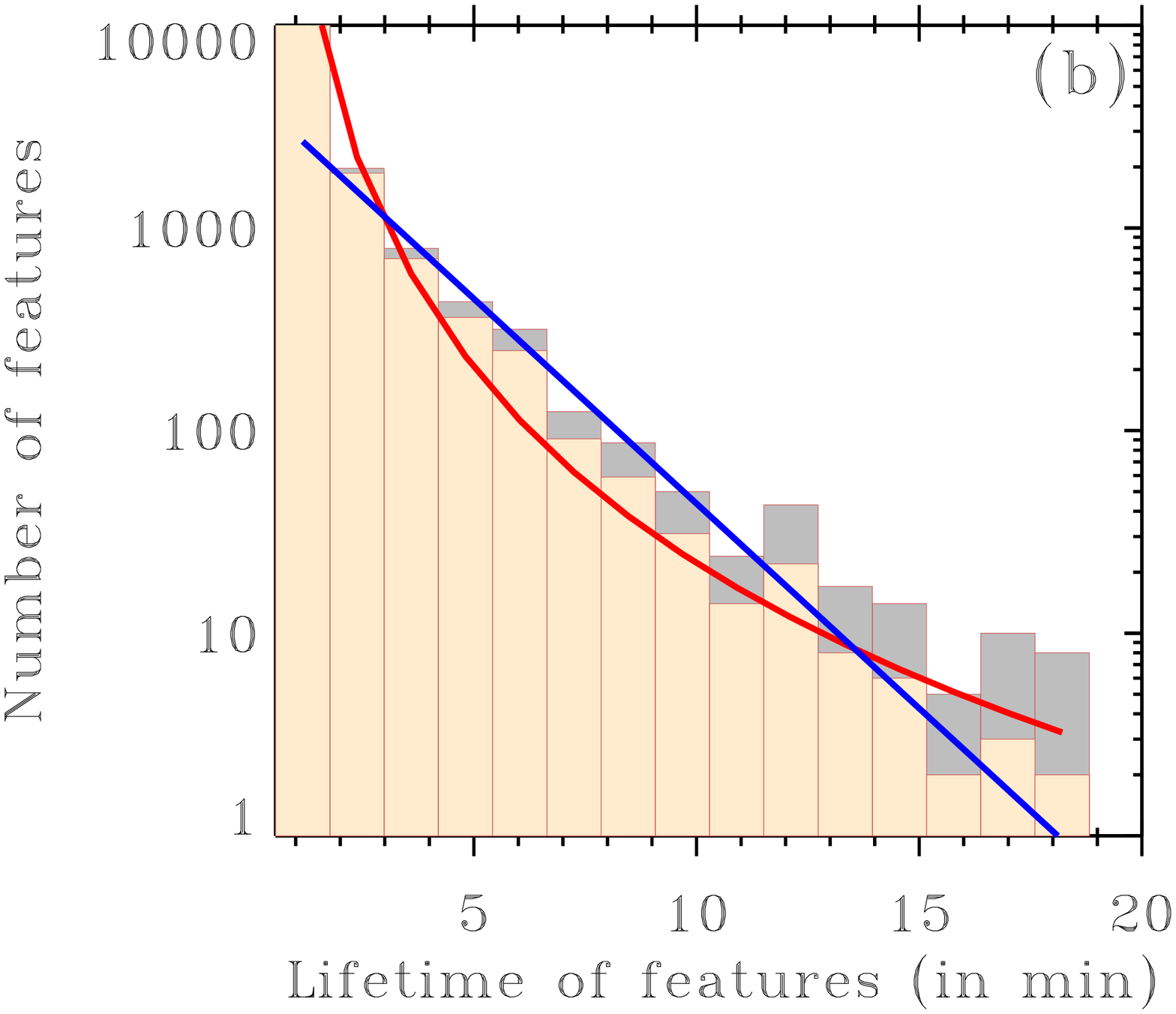}
\caption{Histograms of lifetimes of the magnetic features for the
area-ratio criterion 10:1 (panel a)
and 2:1 (panel b). The red and the blue lines represent the power-law and 
the exponential fits respectively. The gray shaded histograms represent the
lifetimes with a correction factor (see Section~\ref{lifetime}) and the yellow shaded 
histograms represent the lifetimes without any correction.
}
\label{fig11}
\end{figure*}

In Fig.~\ref{fig11} we plot histograms of lifetimes of the features, where the lifetime is 
determined as described in Section~\ref{def}. A correction factor was proposed 
by \citet[][]{danilovicetal2010} to avoid underestimating the lifetimes for the 
same observations, as the duration of the time-series is itself comparable to the 
lifetimes of longer-lived features. The suggested weight to be multiplied to each 
bin of the lifetime distribution is $(n-2)/(n-1-m)$, where $m$ is the number of 
frames a feature lives, and $n$ is the total number of frames. We show lifetime histograms
before (yellow-shaded) and after (gray-shaded) applying the correction factor. The plotted 
histograms are for the two extreme ratios of the areas of the children/parents, 
10:1 and 2:1 respectively. Nearly $39$\,\% and $36$\,\% of the features for the 10:1 and 
2:1 cases respectively, in the lifetime histograms live for just one time step.
The more restrictive the area-ratio criterion restricting the death of features, 
the larger the number of long lived features, since a higher 
number of splitting and merging events are not counted as contributing to the 
birth or death of the larger feature (although for the smaller 
features they are involved in these events). 

The mean and median lifetimes are $1.4$ and $1.1$ minutes ($85$ 
and $66$ seconds) respectively for the 10:1 case and are $1.6$ and $1.1$ minutes ($96$   
and $66$ seconds) respectively for the 2:1 case. 
Similar lifetimes were also reported by \citet[][]{zhouetal2010}.
These small mean and median 
lifetimes are clearly dominated by the large number of features
that live for just one time step. Thus, the mean lifetimes we find are much shorter 
than the lifetimes of the bright points found by \citet[][]{jafarzadehetal2013} using the same 
Sunrise data set. 

We fitted an exponential distribution of the form $N(\tau)=a\,e^{b\tau}$ and a 
power law distribution of the form $N(\tau)=a\,b^\tau$ to the lifetime histograms,
where $N(\tau)$ is the number of events with a lifetime between $\tau$ and 
$\tau+\delta \tau$. In the case of the exponential distribution we refer to the 
parameter $b$ as the `exponential fit index' and in the case of the power-law distribution
we refer to $b$ as the `power-law index'. 
The fit parameters for the two types of distributions are tabulated in 
Table~\ref{table_7}. It reveals that the $\chi^2$ of the fit is smaller for the
exponential distribution in the 10:1 case than that for the power-law distribution 
and is smaller for the power-law distribution than that for the exponential distribution 
in the 2:1 case. On the one hand the strict case of 2:1 area-ratio criterion supports the 
conclusion of \citet[][]{lambetal2013} that a power law with an index of $-2.6$
describes best their lifetime distribution. On the other hand the liberal 10:1 
area-ratio criterion supports the exponential distribution proposed by \citet[][]{zhouetal2010}.
The power law index that we find is in the range $-3.2$ to $-3.7$ and the exponential
fit index is in the range of $-0.5$ to $-0.7$ for the different area-ratio criteria 
(see Table~\ref{table_7}). The different power law indices obtained from our study 
and those by \citet[][]{lambetal2013} can be attributed to a variety of 
reasons (see discussions below). 

\begin{figure*}
\includegraphics[scale=0.4]{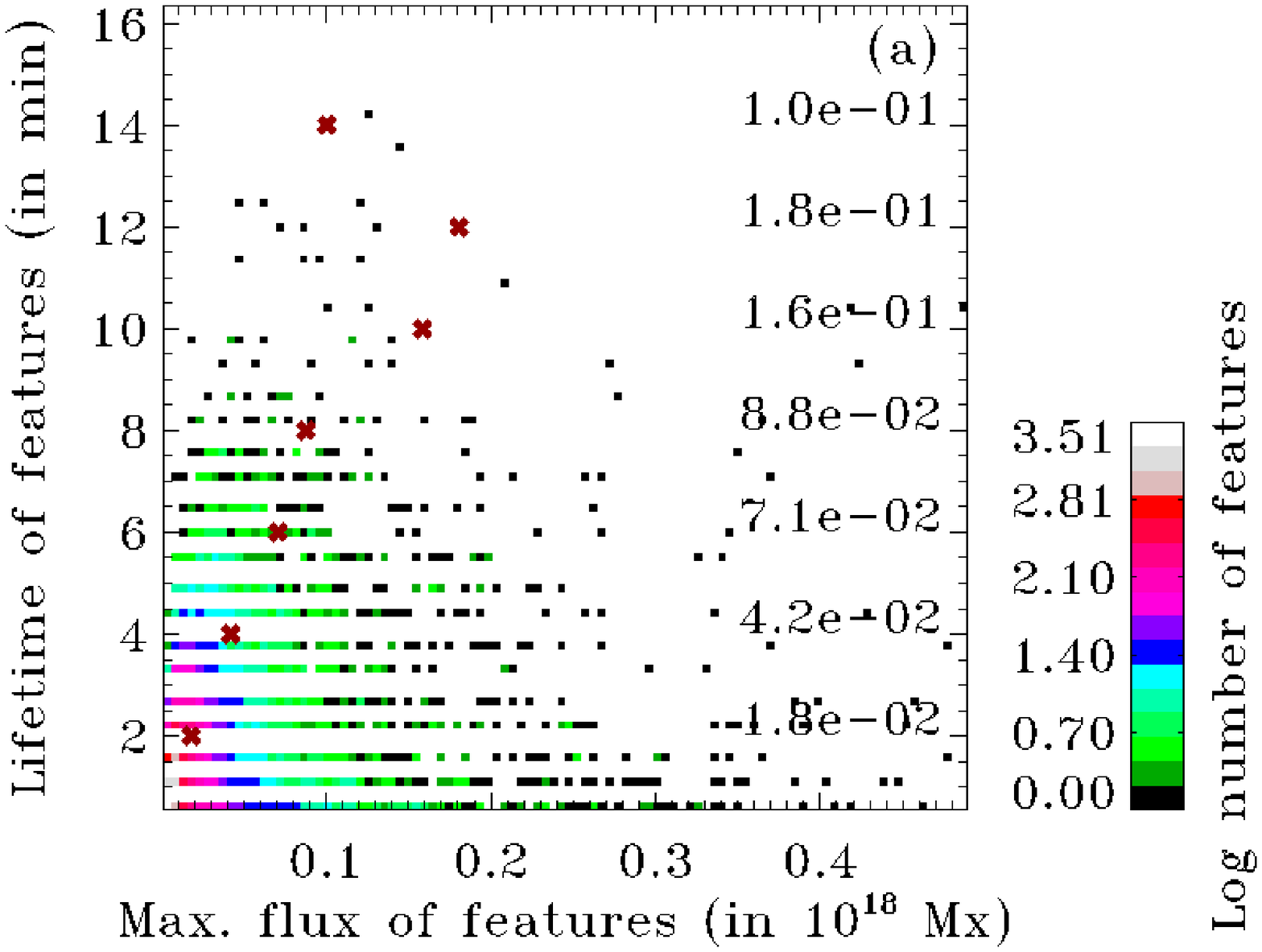}
\hfill
\includegraphics[scale=0.4]{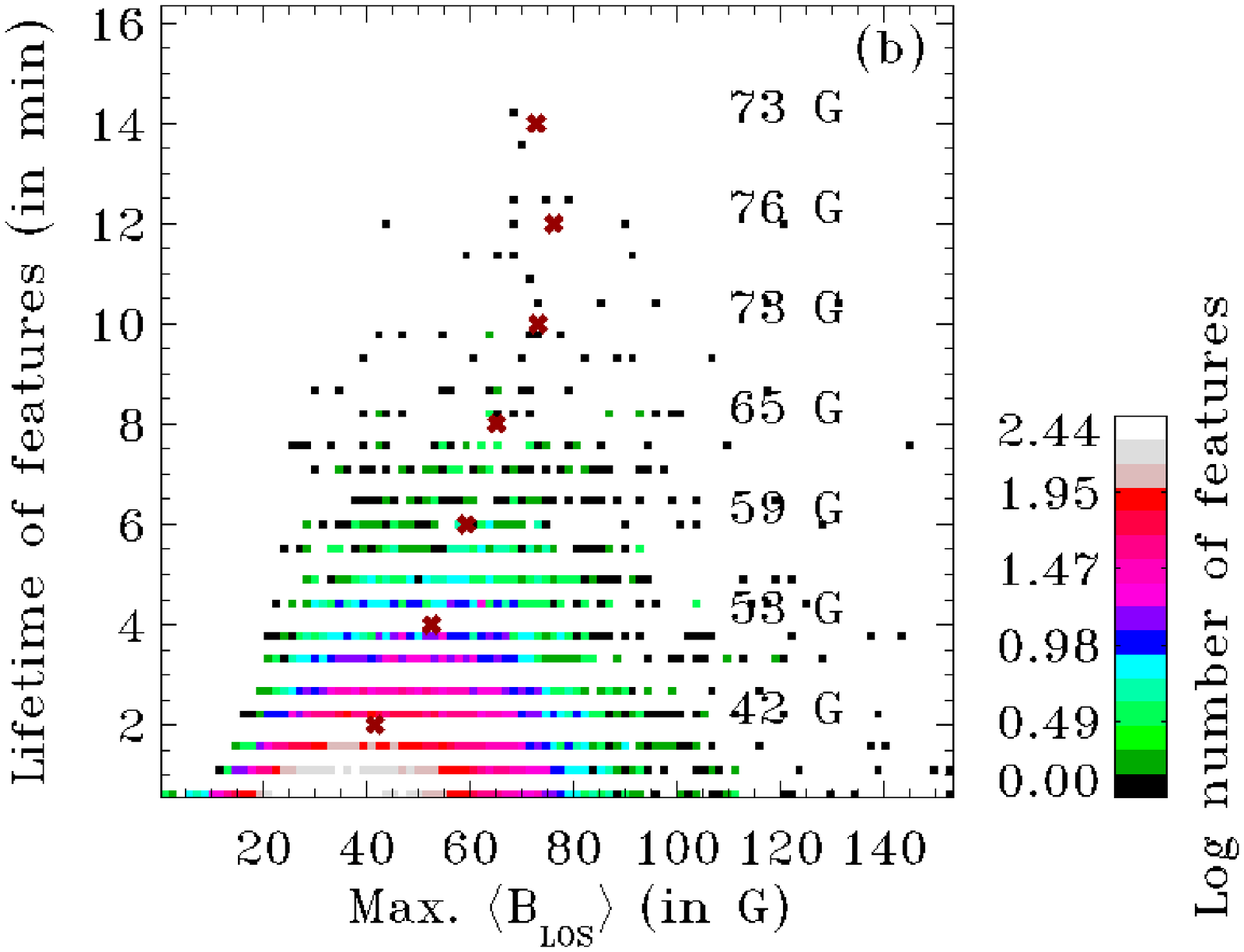}
\caption{Lifetimes of features for an area ratio criterion 5:1 vs. 
maximum fluxes (panel a) and maximum, average magnetic field strength values from 
COG technique (panel b). Here maximum refers to the maximum value of the magnetic 
flux or the field strength, gained during the entire lifetime of the 
feature. In panel (a) the range of plotted 
maximum flux of feature is limited to $5\times10^{17}$ Mx to enhance the 
visibility of flux-lifetime relationship. The over-plotted red diamond symbols represent the 
average value of the maximum flux of features at each lifetime (see main text for details).
The maximum $\la B_{\rm{LOS}}\ra$ averaged over each lifetime bin 
are displayed in the figure and are represented by over-plotted red 
diamond symbols.}
\label{fig-scat3}
\end{figure*}

In Fig.~\ref{fig-scat3} (a) we present a 2D histogram of the number of features
in the plane spanned by the feature lifetime and the maximum magnetic flux achieved by a 
feature during it's lifetime. The plot is for area-ratio restricted to 
a maximum of 5:1. Clearly, the features with a very short lifetime (just 1 or at most a few frames) 
cover the full range of fluxes of features considered here, whereas the long-lived 
features have small flux. 
Interestingly, the region ``high flux long-lived features'' is very sparsely populated. 
For a better visibility of this, the range of plotted
maximum flux of feature is limited to $5\times10^{17}$ Mx.
Although we do not plot a few features that have larger flux, their number is negligible
compared with all the features (36 compared with a total of 15691 features).
The over-plotted red diamond symbols represent the average
value of the maximum flux of feature at each lifetime. This average
includes also the features that lie outside the frame. It is now
clearly visible from these diamond symbols that the maximum flux per feature on average
although increases with lifetime for features living less than 12 minutes, it decreases 
with increasing lifetime for features living longer than 12 minutes.
Since the area and the flux are linearly related, this means that big long-lived features
are also rare. This is somewhat in contrast to the general expectation that larger magnetic 
features live longer, as is the case for sunspots at the extreme end of the area range. 
These follow the Gnevyshev-Waldmeier rule of \citet[][]{gnevyshev1938} and \citet[][]{waldmeier1955} 
\citep[see also][]{solanki2003}, that gives a 
linear relationship between area and lifetime. Of course, even the largest 
features we see here are smaller than sunspots or pores. 
The absence of ``old big features'' is mainly due to the fact that 
they tend to split and merge with other features rather quickly.

The relationship between lifetimes and LOS magnetic field is plotted in 
Fig.~\ref{fig-scat3} (b). Here the largest LOS magnetic field component, 
spatially averaged over the magnetic feature, reached during the lifetime 
of the feature is displayed. The figure shows that short-lived features 
mostly appear in the magnetic field range of $0$ to $150$\,G. 
There is a weak tendency for the LOS magnetic field to increase with 
lifetime (as can be gleaned from the bin-averaged values given in 
Fig.~\ref{fig-scat3} (b), that are represented using red diamond symbols).

\subsection{Evolution of magnetic features}
\label{evolution}
\begin{figure}
\includegraphics[scale=0.45]{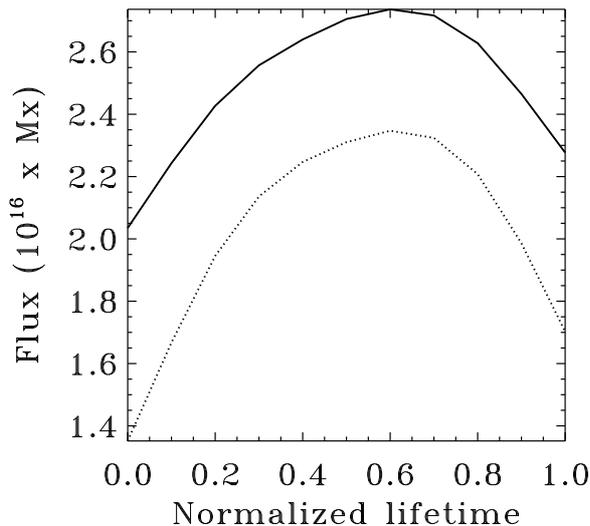}
\hspace{0.5cm}
\caption{Average growth and decay of fluxes of all the features with
area-ratio criterion 10:1 (solid line, averaged over $3632$ features) 
and 2:1 (dotted line, averaged over $3459$ features). These features live
for at least 4 time steps (2 minutes).}
\label{fig-growth-decay-live}
\end{figure}

\begin{figure*}
\sidecaption
\parbox{12cm}{
\includegraphics[width=6cm]{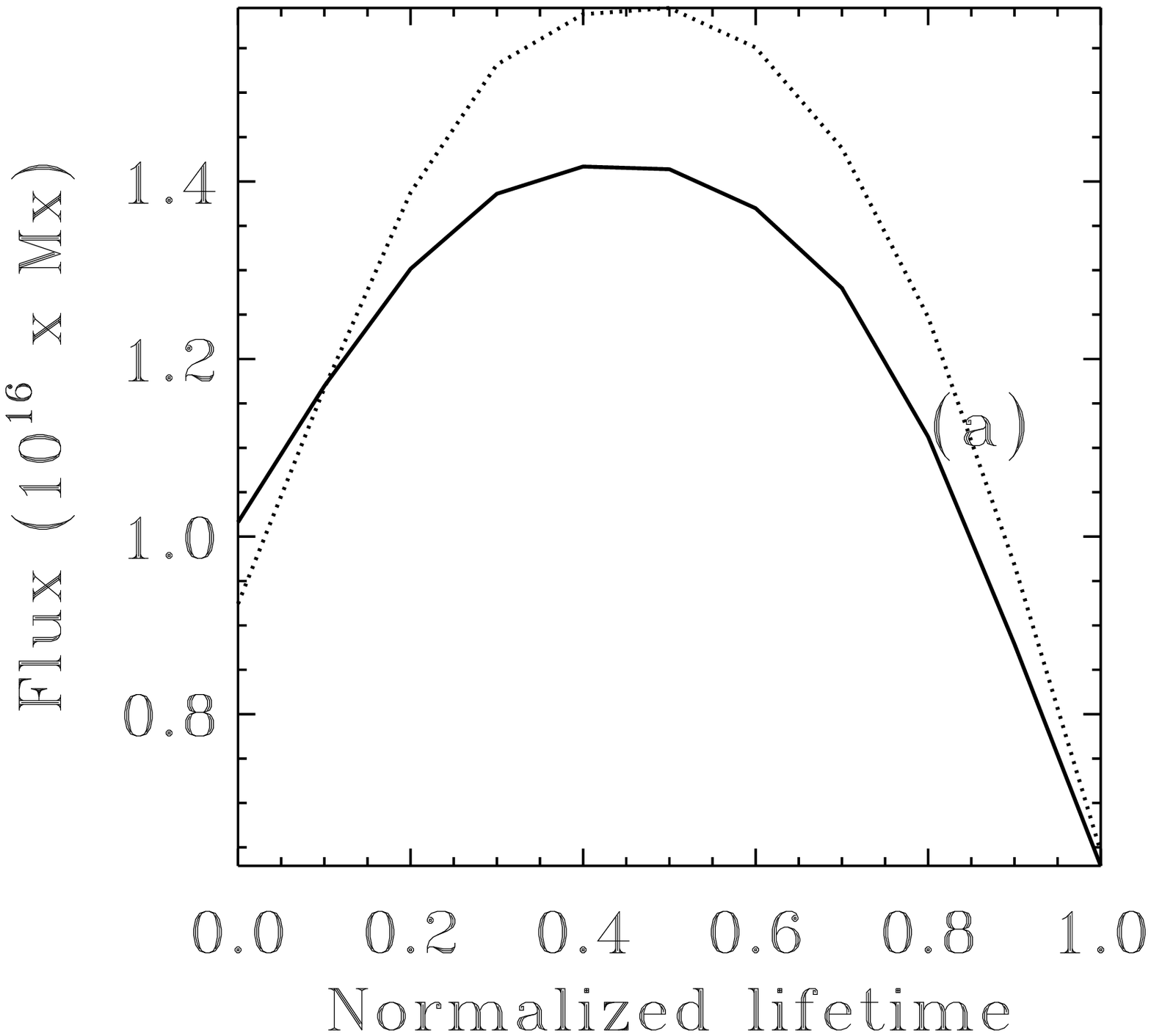}
\hfill
\includegraphics[width=6cm]{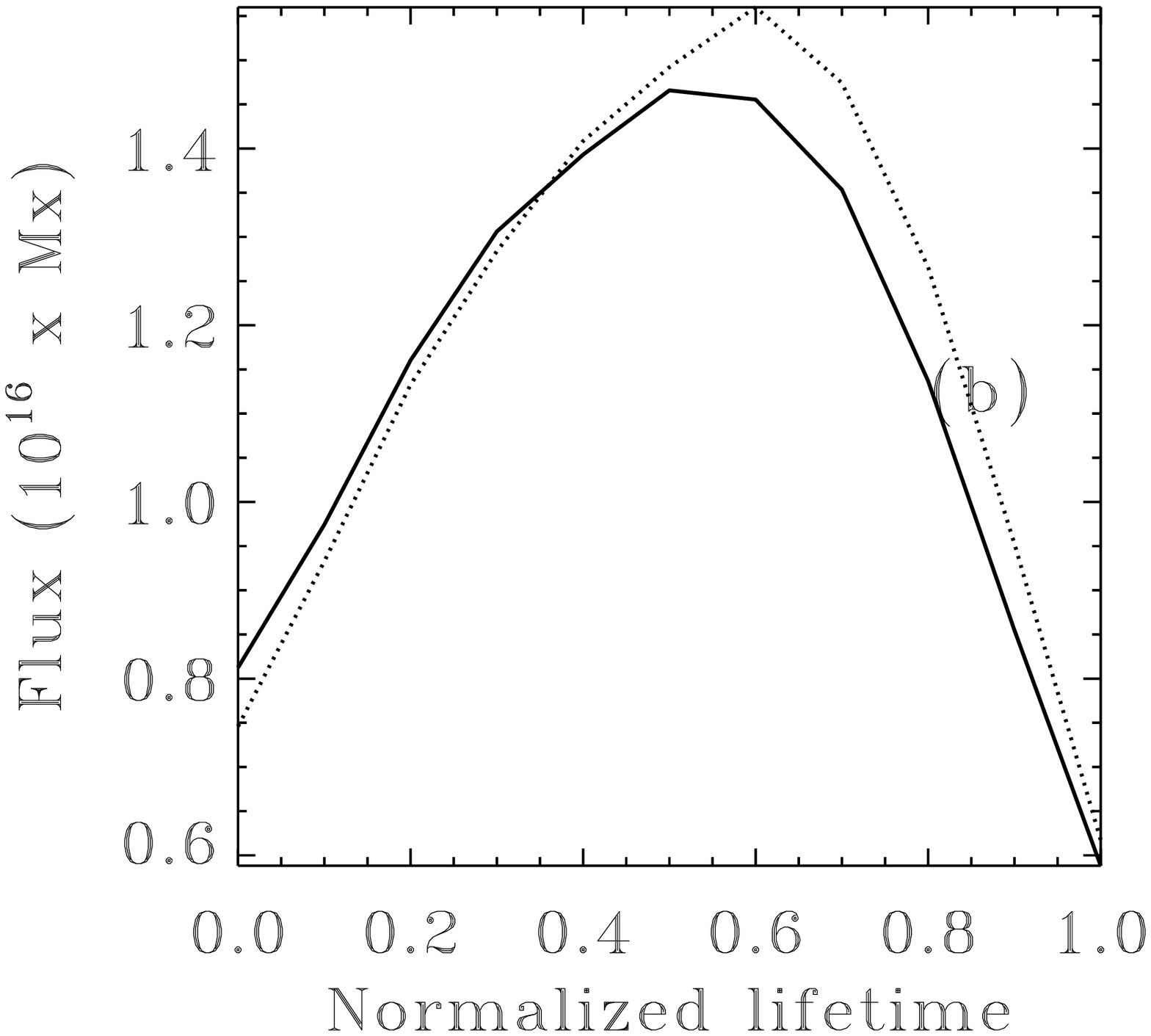}
\includegraphics[width=6cm]{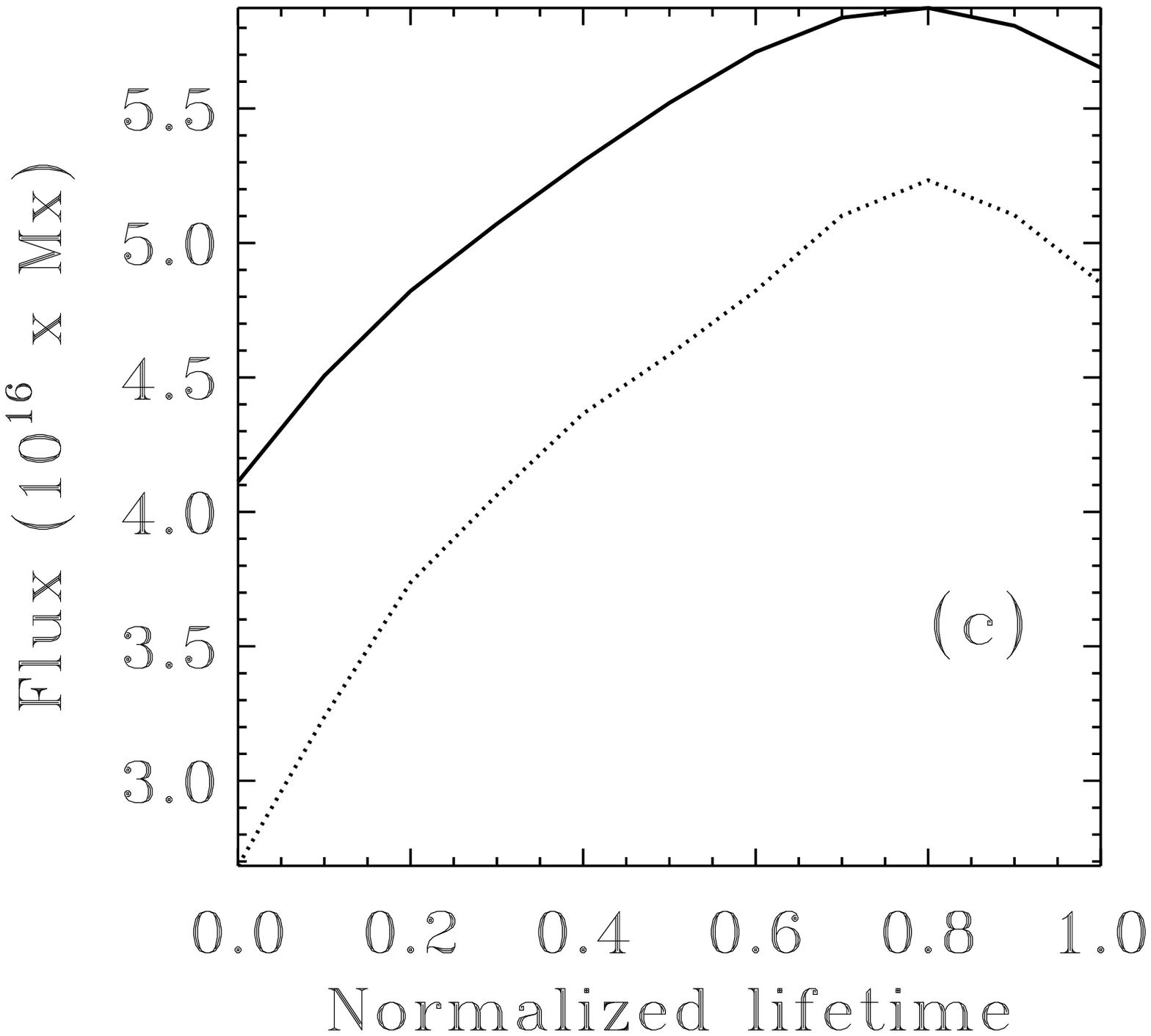}
\hfill
\includegraphics[width=6cm]{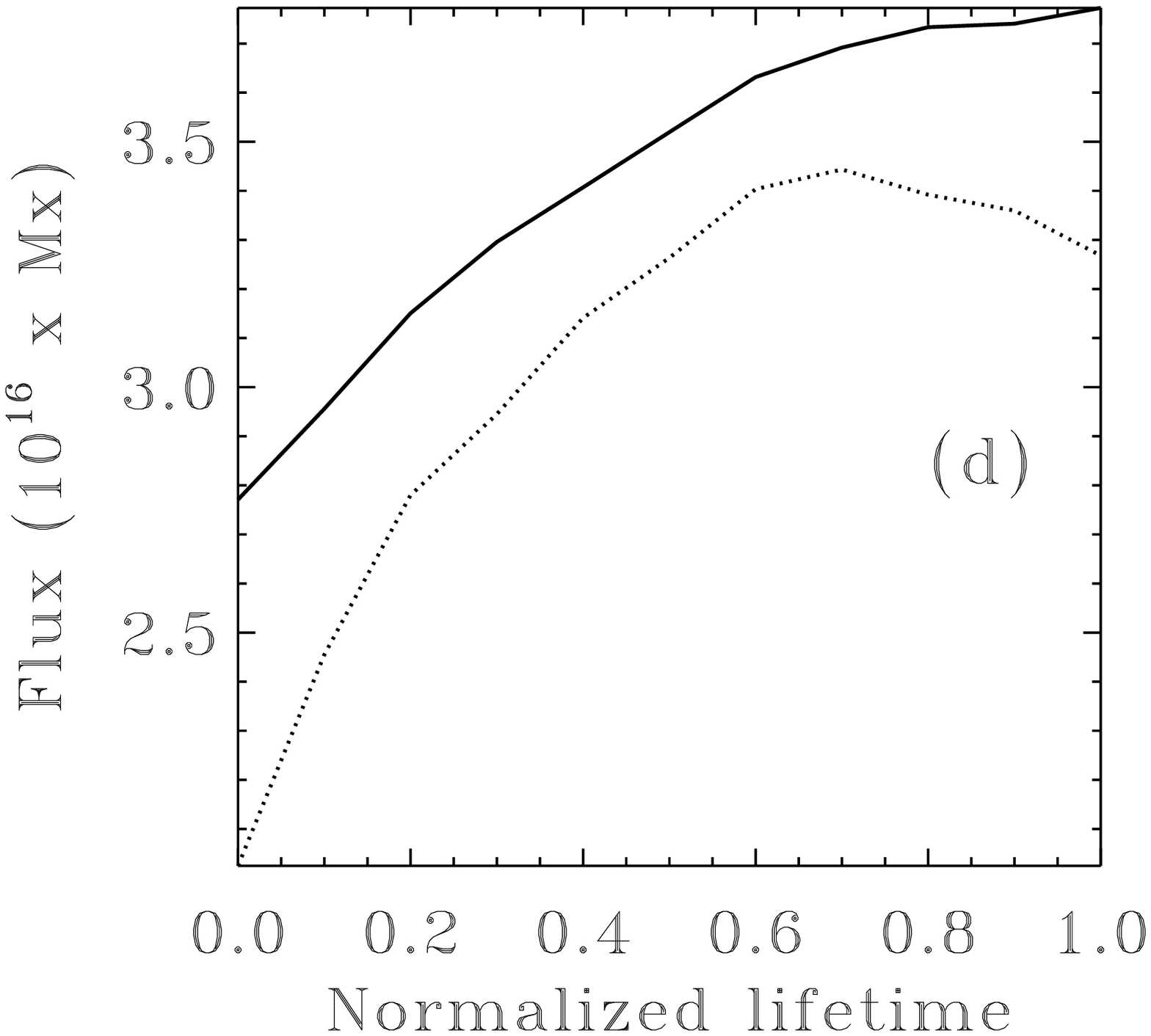}
}
\caption{(a) Same as Fig.~\ref{fig-growth-decay-live}, but for the features 
that die by disappearance for area-ratio criterion 10:1 (solid line, averaged 
over $1994$ features) and 2:1 (dotted line, averaged over $2224$ features). 
(b) Same as (a) for features that die by cancellation (averaged over 
$113$ and $126$ features for 10:1 and 2:1 respectively).
(c) Same as (a) for features that die by splitting (averaged over 
$804$ and $523$ features for 10:1 and 2:1 respectively). (d) Same as (a) for 
features that die by merging (averaged over $687$ and $333$ features for 10:1 
and 2:1 respectively).} 
\label{fig-growth-decay-die}
\end{figure*}

\begin{figure*}
\sidecaption
\parbox{12cm}{
\includegraphics[width=6cm]{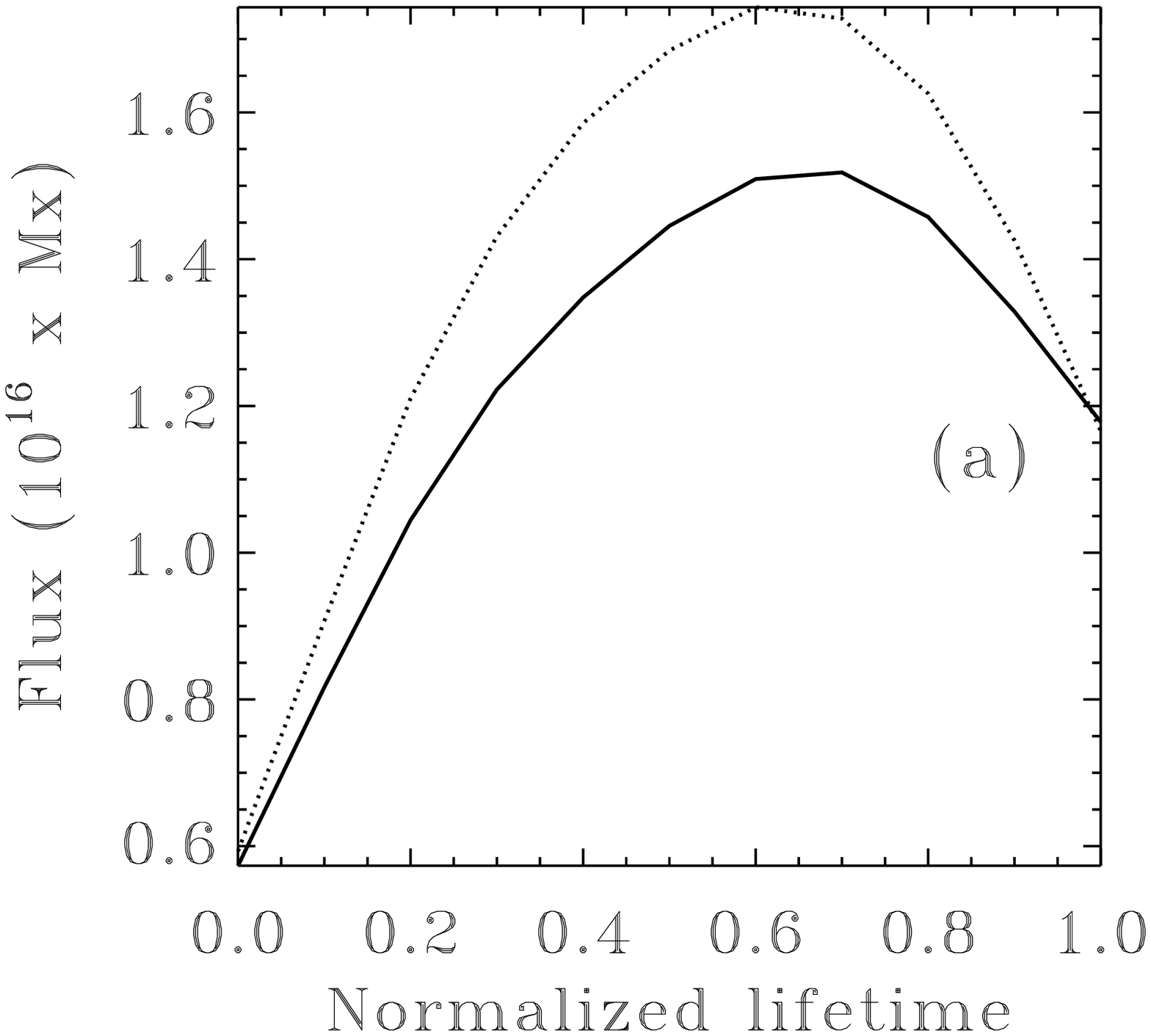}
\includegraphics[width=6cm]{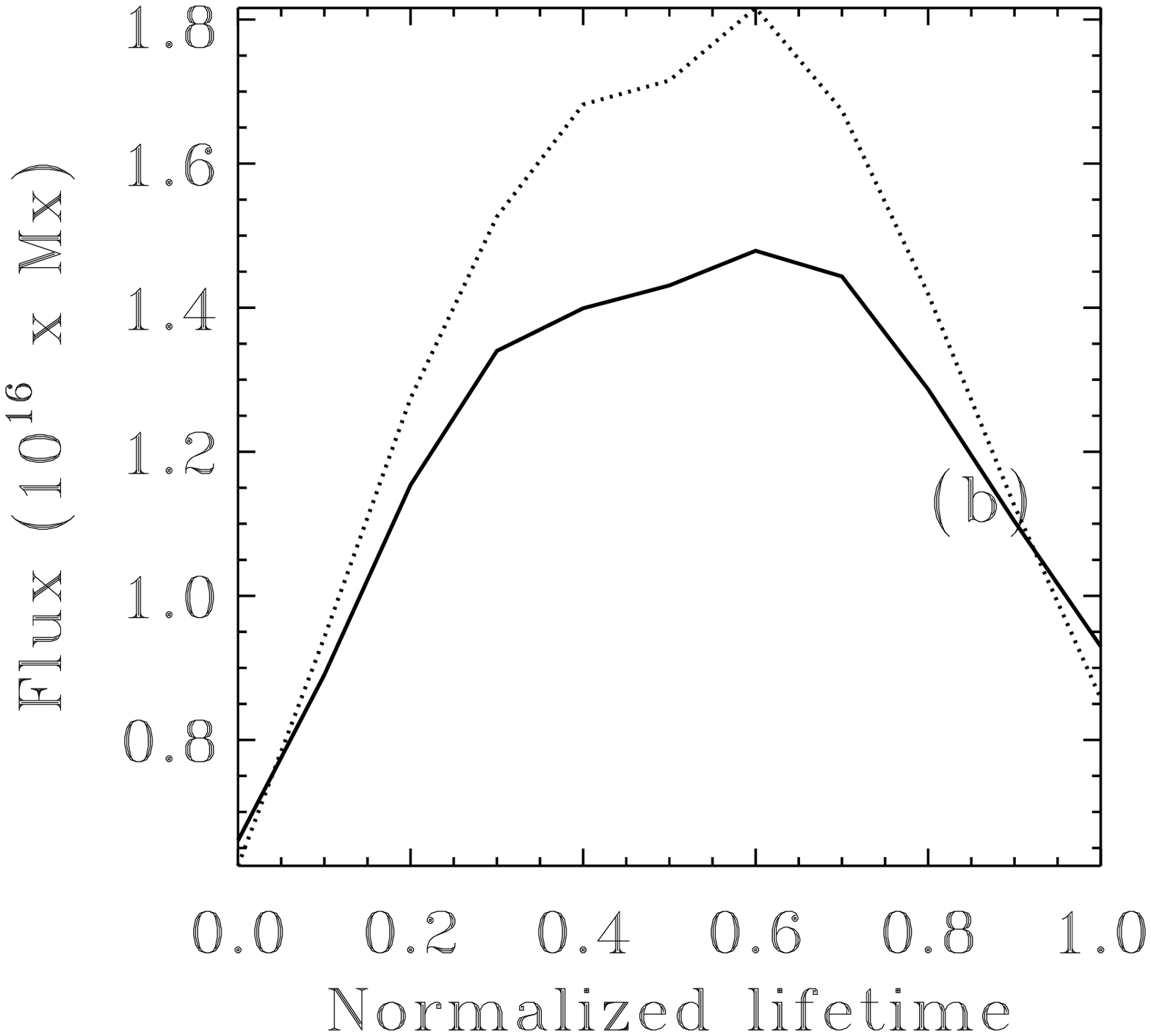}
\includegraphics[width=6cm]{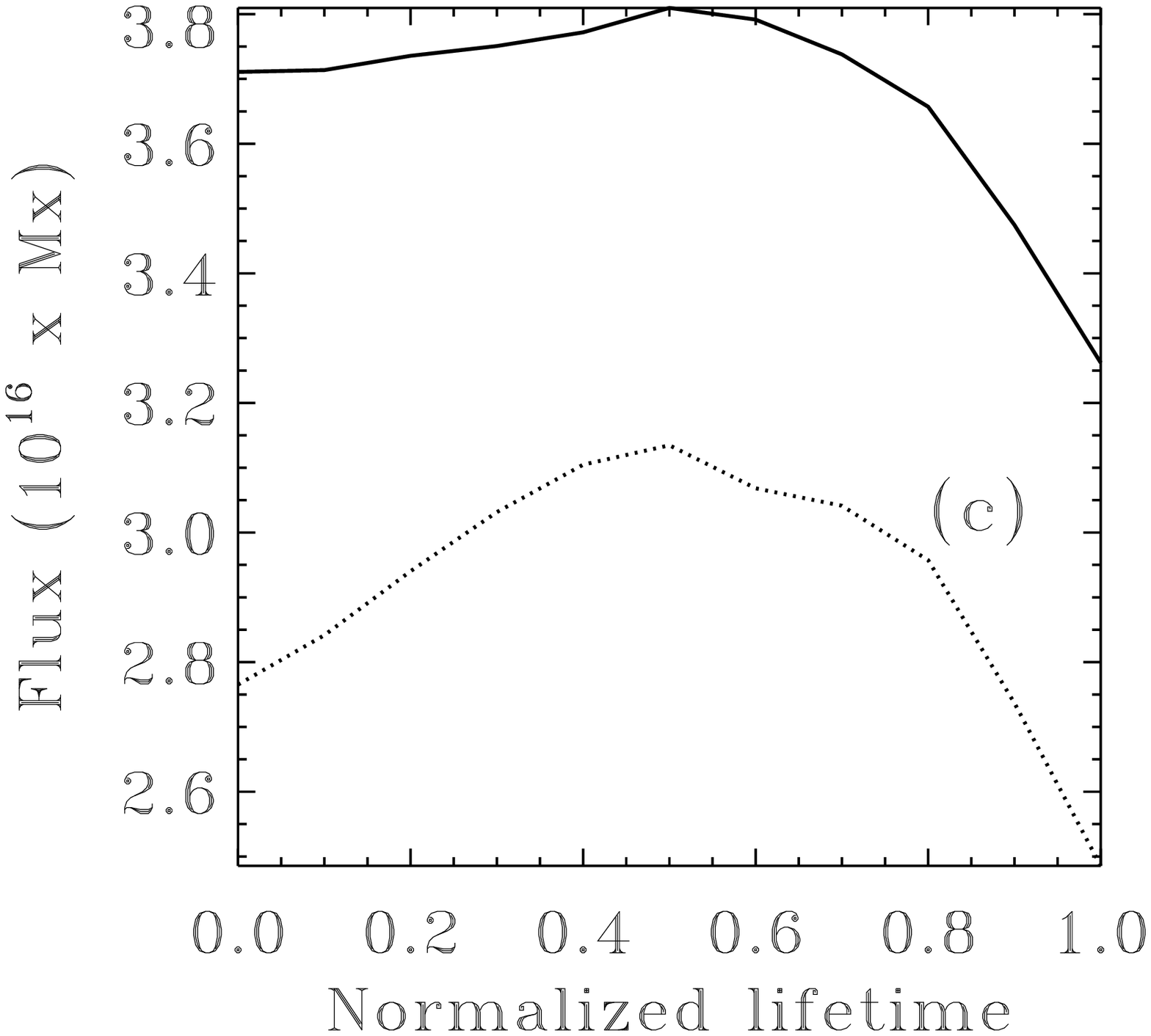}
\includegraphics[width=6cm]{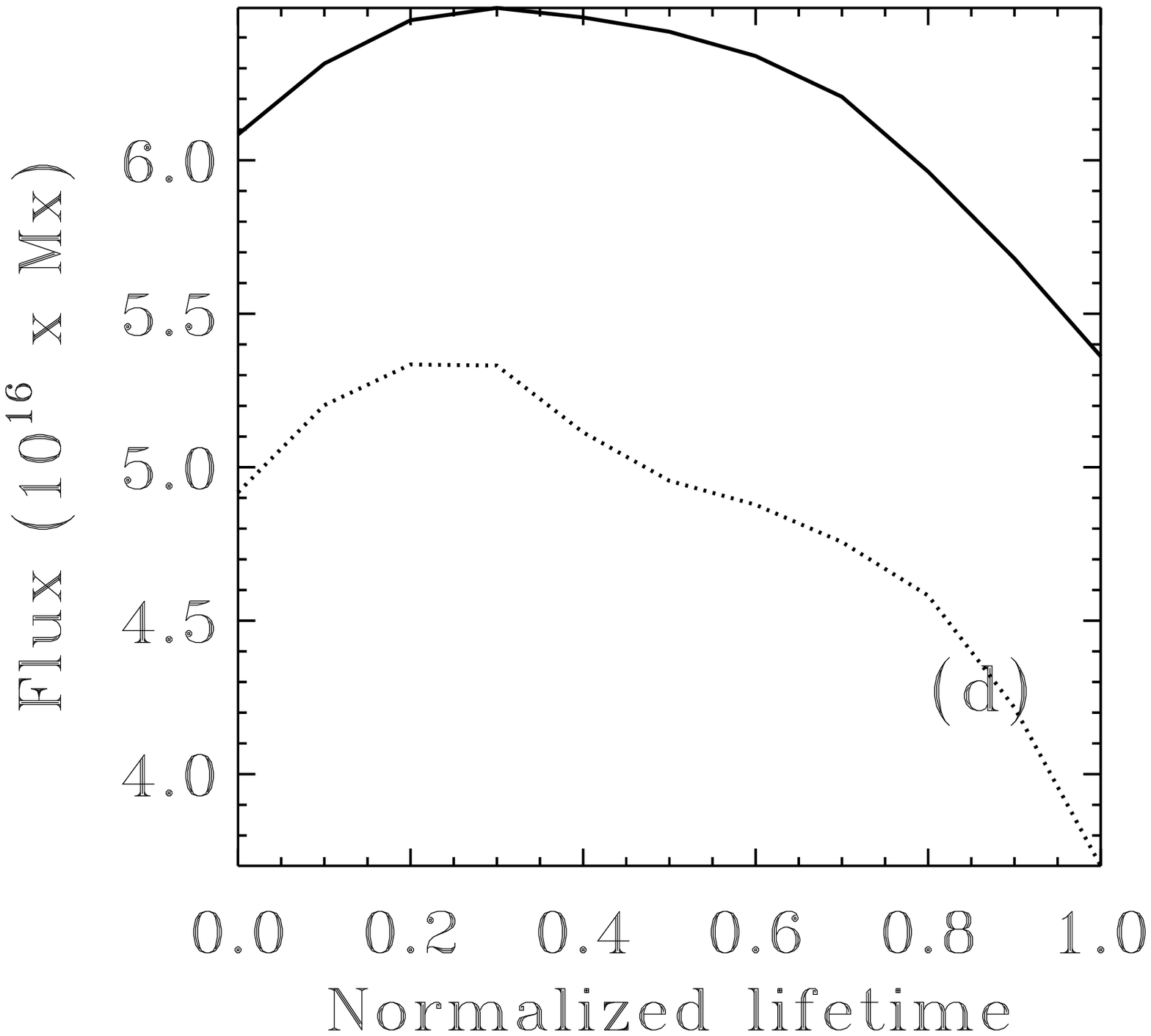}
}
\caption{(a) Same as Fig.~\ref{fig-growth-decay-live}, but for features born by 
appearance for area-ratio criterion 10:1 (solid line, averaged over $2116$ features) and 
2:1 (dotted line, averaged over $2425$ features). (b) Same as (a) 
for features born by emergence (averaged over $64$ and $25$ features for 10:1 and 2:1 
respectively). (c) Same as (a) for features born by splitting (averaged over 
$791$ and $375$ features for 10:1 and 2:1 respectively). (d) Same as (a) for 
features born by merging (averaged over $631$ and $413$ features 
for 10:1 and 2:1 respectively).}
\label{fig-growth-decay-born}
\end{figure*}

We now consider the evolution of the magnetic fluxes of the 
features which live for at least 4 time steps (2 minutes).
In Figs.~\ref{fig-growth-decay-live}--\ref{fig-growth-decay-born} 
we plot feature-averaged fluxes as a function of normalized lifetimes.
Fig.~\ref{fig-growth-decay-live} shows growth and decay of all features
that are born and died during the analyzed time series. 
On average, the fluxes in the features increase by $20-30$\,\% 
after their birth, reach a maximum after around $2/3$ of their lifetime 
lies behind them and then decrease again by $10-20$\,\% until their death. 
Consequently, the variation over the lifetime is relatively small. 
Interestingly, the features have on average a higher magnetic flux 
at the end of their lifetime than at the beginning. The dotted and solid 
curves are for the 2:1 and 10:1 area-ratio criterion, respectively. 
Only the evolution of magnetic flux is plotted since the area displays 
a nearly identical evolution.

The flux evolution of features depends sensitively on the method of 
birth and/or death, as is illustrated in Figs.~\ref{fig-growth-decay-die} and 
\ref{fig-growth-decay-born}. In Figs.~\ref{fig-growth-decay-die} (a) and 
\ref{fig-growth-decay-die} (b) the magnetic flux evolution of features that undergo 
death by disappearance and cancellation is depicted, respectively. The magnetic 
flux of such features peaks nearly in the middle of their lifetime, and 
drops rapidly after that, reaching a value that is on average half of their peak flux.  
Note that, although the features considered here have died by disappearance and 
cancellation, these features could be born by any type of birth event. Therefore
the curves depend on the area-ratio criterion. 
In Figs.~\ref{fig-growth-decay-die} (c) and \ref{fig-growth-decay-die} (d) 
we display the evolution of features that die only due to splitting and merging 
respectively. These display nearly the opposite evolution, with the flux growing 
(by a factor of roughly $1.3-1.8$) until nearly the time of death.

In Figs.~\ref{fig-growth-decay-born} (a) and \ref{fig-growth-decay-born} (b) 
the evolution of features that are born by appearance 
and emergence are plotted respectively, while Figs.~\ref{fig-growth-decay-born} (c) and
\ref{fig-growth-decay-born} (d) show the evolution of features
that are born exclusively by splitting and merging, respectively.

The growth and decay of magnetic flux over a feature's lifetime often behaves 
differently for the two extreme area-ratio criteria. 
In most cases with sufficiently large samples of features the two curves display 
a qualitatively similar shape, but have different magnitudes, corresponding to 
different amounts of magnetic flux per considered feature. The flux per feature is 
larger for the 10:1 area ratio criterion in the cases of the features dying by
splitting and merging. This is because, in the liberal 10:1 case, a big feature 
dies even if only a small part of it breaks off, whereas in 
the 2:1 case such a big feature would continue to live on. Hence, a feature that is 
counted only once in the 2:1 is counted multiple times in the 10:1 case, so that the 
average flux per feature is increased. 

Conversely, for the features that died by disappearance and cancellation, 
the average flux per feature is larger for the 2:1 area-ratio criterion. This 
is because, in the 10:1 case, most big features die by splitting and merging 
and the remaining smaller features, those which die by disappearance or cancellation, 
begin their journey with a smaller flux and have relatively shorter life times. 
In the 2:1 case, far fewer big features die by splitting and merging and therefore 
they have higher initial flux and longer life times (see also Fig.~\ref{fig11}). 
Since here we consider only features which live at least for 4 time
frames, the above argument leads to a larger flux per feature for the 2:1 case. 
However note that the initial flux of the features in Fig.~\ref{fig-growth-decay-die} 
is the average of the initial fluxes of the features that are born by any birth event. 
Therefore the initial fluxes behave differently, in comparison to the behavior of
the average flux per feature discussed above.

The features that die by cancellation and disappearance show smaller fluxes 
than the features that die by splitting and merging. 
At their death, the flux of disappearing and cancelling features is smaller than 
the flux at their births, which is not surprising, since these processes 
either remove flux from the solar surface, or redistribute it in such a way that 
it gets hidden from observations of Stokes $V$. Possible ways of flux removal are 
by retraction of a small magnetic loop, often preceded by magnetic reconnection, 
or the ejection of a $U$-loop (e.g. produced by reconnection below the solar surface), 
or an $O$-loop, or the dissipation of magnetic flux structured at very small scales. 
Flux at the solar surface can be hidden by, e.g., diffusion of the flux until the Stokes 
$V$ signal it produces drops below the noise level, or by the splitting of small 
magnetic features, so that the resulting features are below the flux limit, by 
bringing opposite polarity features close to each other (opposite polarities within 
a resolution element lead to the cancellation of Stokes $V$ signals), or increased 
inclination of the magnetic field. The last of these processes works even when 
considering the full Stokes vector because for a given noise level in Stokes 
$Q$, $U$, $V$, it requires nearly an order of magnitude larger horizontal magnetic 
flux (in the line formation layer) per pixel for a $Q$ or $U$ signal to be visible 
above the noise than vertical magnetic flux for it to be detectable in Stokes $V$
\citep[see also][and the references cited therein]{lambetal2013}. 

Features that die due to splitting and merging reach their largest 
fluxes in the second half of their lifetimes. On average, they have higher fluxes at the time 
of their death than at birth. Clearly, it is the biggest features that tend to split.
This is very similar to the behavior of granules, where it is
also the larger granules that are likely to split \citep[see][]{hirzbergeretal99}, 
although the physical reasons are likely to be different. We note that at 
least partly the tendency for the big features to split is a result of a 
bias: if a feature with an area of less than 10 pixels splits, at least one 
of the children will be below the minimum size requirement of 5 pixels, 
so that no splitting will be seen. Indeed, if the area is 8 pixels or less, 
the feature might even ``disappear'' according to our criteria.

In Figs.~\ref{fig-growth-decay-born} (a)--\ref{fig-growth-decay-born} (d)
the evolution of features classified according to their birth type
are shown, which can be explained using similar arguments used for understanding
the evolution of the features classified according to death, discussed above.


\section{Conclusions}
\label{conclusions}
In this paper we present a newly developed feature tracking algorithm and use
it to study the evolution of small-scale magnetic features present in quiet-Sun 
high spatial and temporal resolution observations made by the IMaX 
instrument on the Sunrise balloon-borne observatory. We use
the COG technique of \citet[][]{reesandsemel1979} for deducing
the LOS components of the magnetic field, from which we determine
the magnetic flux.
We apply the technique to Stokes $V$ signals spatially averaged over the magnetic
features. The feature-averaged magnetic field strength values thus obtained 
range between $50-150$\,G. Note, however, that maximum field strengths can 
reach kG values as shown in Fig.~\ref{fig-flux-calib1} even when 
assuming that the features are spatially resolved 
\citep[see also][]{laggetal2010,requereyetal2014,riethmulleretal2014}.

We focus our attention on the processes by which these features are born or
die, namely birth by uni-polar appearance, splitting, merging, bi-polar emergence,
death by uni-polar disappearance, splitting, merging and by cancellation of opposite
polarities.

Qualitatively our result agrees with other recent studies 
\citep[see e.g.,][]{lambetal2008,lambetal2013}. In summary, bi-polar 
emergence events are rare compared to the birth of the magnetic features 
by uni-polar events, emergence being responsible for only $0.3-0.5$\,\%
of the total instantaneous flux and only $0.5-1$\,\% of the total maximum flux 
in newborn features. On the other hand, appearances 
account for $12$\,\% of the total instantaneous flux and $19.1$\,\% of the 
total maximum flux, which are much higher than the fluxes in emergence events.
The remaining flux is distributed in splitting, merging and merging-into birth events.
When only appearance and emergence are considered as birth events (leaving out splitting 
and merging), flux gained in appearance accounts for $92.4$\,\% of the total 
instantaneous flux, and $91$\,\% of the total maximum flux of the 
new flux at the solar surface. The remaining flux is
is divided among the features involved in time-symmetric and time-asymmetric emergences.

Death by cancellation corresponds to $1.7$\,\% of the total instantaneous 
flux and $2.9$\,\% of the total maximum flux for the area ratio criterion of 10:1,
while disappearance corresponds to a higher flux loss, which is 
$10.3$\,\% of the total instantaneous flux and $16.2$\,\% of the total 
maximum flux. The remaining flux is distributed in other uni-polar death types, namely, 
splitting, merging and splitting-off events. When only disappearance and 
cancellation are considered as death events (leaving out splitting and merging), 
cancellation leads to $14$\,\% of the total instantaneous flux and $15$\,\% of the 
total maximum flux. The features that survive in semi and partial cancellation
also remove flux. After including this, the flux lost in 
disappearance is $59$\,\%, flux lost in cancellation leading to death of features 
and in those surviving cancellation is respectively, $10$\,\% and $31$\,\%.

The fact that a large number of features simply appear and disappear, without 
any clear sign of an opposite polarity, is partly due to the limited spatial 
resolution of the Sunrise/IMaX observations (although higher than that of the data 
used for previous such studies). E.g., if two features with sizes below the 
threshold of $5$ pixels each merge, then the resulting feature can be larger 
than this threshold. It will then be seen to have simply appeared. Similarly, 
a feature of $5-8$ pixels can either split into two features that are below 
the resolution threshold of $5$ pixels so that it disappears, or equivalently, 
it can partially cancel with a feature just below the $5$ pixel threshold 
and also disappear. Such interactions, as well as those with features hidden 
because their Stokes $V$ signals are not sufficiently above the noise, will go undetected.

The lifetime of the magnetic features follows a distribution whose exact 
shape is not completely clear. Thus it is best described by a power law
for the 2:1 area-ratio criterion and an exponential function for the 10:1 
area-ratio criterion. These functions agree with
those suggested by \citet[][Hinode data]{lambetal2013} for the 2:1 area-ratio 
while for the 10:1 case the best-fit function agrees with 
\citet[][Hinode data]{zhouetal2010}. The power-law fit index ranges from $-3.2$ to 
$-3.7$ and the exponential fit index ranges between $-0.5$ to $-0.7$.

The area, instantaneous flux and maximum flux of the magnetic features also 
follow power-law-type distributions with exponents of $-2.25$, $-1.85$ and 
$-1.78$, respectively. They are in agreement with previous studies, 
by \citet[][area distribution, Hinode data]{buehleretal2013} and 
\citet[][flux distribution, several data sets of last two decades]{parnelletal09}, 
\citet[][flux distribution, Hinode data]{iidaetal2012}. However we find that the power-law
indices sensitively depend on the smallest flux per feature chosen and, even more strongly, 
on the type of binning. 
Because of the higher spatial resolution of Sunrise/IMaX instrument we can detect 
features of flux as low as $9\times10^{14}$ Mx which is nearly an order of magnitude
lower than the lowest flux detected in previous studies using Hinode data.

We also studied the growth and decay of the features. This study suggests that the features
that appear, disappear, emerge and cancel carry low amounts of magnetic flux when compared
to the fluxes in the features that split and merge. This suggests that the largest features 
are those who tend to split, followed by those who merge, and finally those who disappear 
and cancel. 

\begin{table*}
\begin{center}
\caption{Parameters of fits to flux distributions in Figs.~\ref{fig5a} and \ref{fig5b}.}
\vspace{0.6cm}
\label{table_1}
\begin{tabular}{crrrr}
\hline
 & \multicolumn{2}{c}{Fit parameters for Fig~\ref{fig5a}} \\
\hline
Type & Instantaneous flux & Maximum flux \\
\hline
Minimum value of flux  & $8.8\times10^{14}$ Mx & $9\times10^{14}$ Mx\\
Power law index  &-1.85  & -1.78 \\
$\chi^2$ of the fit & 63.7 &  28.3 \\
\hline
 & \multicolumn{2}{c}{Fit parameters for Fig~\ref{fig5b}} \\
\hline
Type & Instantaneous flux & Maximum flux \\
\hline
Minimum value of flux  & $4.7\times10^{15}$ Mx & $4.8\times10^{15}$ Mx\\
Power law index  & -0.97 (Blue) & -1.04 (Blue)\\
$\chi^2$ of the fit & 2.03 & 2.16 \\
\hline
Minimum value of flux  & $2\times10^{16}$ Mx & $2\times10^{16}$ Mx\\
Power law index  & -1.16 (Red) & -1.23 (Red)\\
$\chi^2$ of the fit & 0.69 & 0.93 \\
\hline
Minimum value of flux  & $8\times10^{16}$ Mx & $8\times10^{16}$ Mx\\
Power law index  & -1.19 (Purple) & -1.33 (Purple)\\
$\chi^2$ of the fit & 0.65 & 0.77 \\
\hline
Minimum value of flux  & $2\times10^{17}$ Mx & $2\times10^{17}$ Mx\\
Power law index  & -1.03 (Green) & -1.26 (Green)\\
$\chi^2$ of the fit & 0.52 & 0.74 \\
\hline
\end{tabular}
\end{center}
\end{table*}

\begin{table*}
\begin{center}
\caption{Number of features born as well as the number that died due to different processes, as listed
in the first column of the table. The following columns give the number of features
involved in birth and death events of a certain type, as well as their percentage contribution 
to all features that are born, or have died, respectively. The individual
columns differ in the applied area-ratios of splitting and merging features or
flux ratios of emerging features (see text for details).}
\vspace{0.6cm}
\label{table_2}
\begin{tabular}{crrrr}
\hline\hline
 & \multicolumn{4}{c}{Area-ratio criteria} \\
\hline\hline
Type of birth & 10:1 & 5:1 & 3:1 & 2:1 \\
\hline
Appearance & 8728 &  8834 &  8922 & 8984 \\
 & (48\,\%) & (52\,\%) &  (55\,\%) &  (59\,\%)  \\
\hline
Splitting & 6072 & 4879 & 3757 & 2505 \\
 & (34\,\%) & (28\,\%) & (23\,\%) &  (17\,\%)\\
\hline
Merging & 2226 & 1846  & 1432 & 965 \\
  & (12\,\%) & (11\,\%) & (9\,\%) &  (6\,\%) \\
\hline
Splitting-off & 646 & 1287 & 1887 &  2561 \\
  & (4\,\%) & (7.5\,\%) & (12\,\%) & (17\,\%) \\
\hline
Emerging & 365 & 259  & 171 &  109\\
  & (2\,\%) & (1.5\,\%) & (1\,\%) &  (1\,\%) \\
\hline
Alive in first frame & 1019 &1019  & 1019 & 1019 \\
\hline\hline
Total & 19056 & 18124 & 17188 & 16143 \\
\hline\hline
Type of death & 10:1 & 5:1 & 3:1 & 2:1 \\
\hline
Disappearance & 7533  & 7533  & 7533 & 7533 \\
  & (42.7\,\%) & (44.8\,\%) & (47.3\,\%) & (50.6\,\%) \\
\hline
Splitting  & 2843 & 2277 & 1753 & 1177 \\
 & (16\,\%) & (14\,\%) &  (11\,\%) & (8\,\%) \\
\hline
Merging  & 5226 & 4232 & 3206 & 2120 \\
 & (29\,\%) & (25\,\%) & (20\,\%) &  (14\,\%) \\
\hline
Merging-into & 627 & 1255  & 1869  & 2486\\
  & (4\,\%) & (7\,\%) & (12\,\%) &  (17\,\%) \\
\hline
Cancelling & 1550 & 1550  & 1550 & 1550 \\
  & (8.7\,\%) & (9.2\,\%) & (9.7\,\%) &  (10.4\,\%) \\
\hline
Alive in last frame & 1277 & 1277 & 1277 & 1277 \\
\hline\hline
Total & 19056 & 18124 & 17188 & 16143 \\
\hline
\hline
\end{tabular}
\end{center}
\end{table*}

\begin{table*}
\begin{center}
\caption{Same as Table~\ref{table_2}, but listing the total magnetic flux (in Mx) of the features 
at the time of their birth or death (first line of each row). Fractional contribution of a certain 
type of birth/death to the total amount of flux in newly born or dying features is expressed in\,\% 
and is given in the second line of each row. The average flux per feature is given in the 
third line of each row.}
\vspace{0.6cm}
\label{table_3}
\begin{tabular}{crrrr}
\hline\hline
 & \multicolumn{4}{c}{Area-ratio criteria} \\
\hline\hline
Type of birth & 10:1 & 5:1 & 3:1 & 2:1 \\
\hline
Appearance: total & 4.7$\times10^{19}$ Mx & 4.8$\times10^{19}$ Mx  & 4.8$\times10^{19}$ Mx & 4.8$\times10^{19}$ Mx \\
(\%) & (12\,\%) & (16.5\,\%) & (21\,\%) & (27\,\%)  \\
Average flux per feature & 5.4$\times10^{15}$ Mx& 5.4$\times10^{15}$ Mx& 5.4$\times10^{15}$ Mx& 5.4$\times10^{15}$ Mx\\
\hline
Splitting: total & 1.8$\times10^{20}$ Mx & 1.2$\times10^{20}$ Mx &  8.2$\times10^{19}$ Mx & 5.3$\times10^{19}$ Mx \\
(\%) & (44.5\,\%) & (41\,\%) & (36\,\%) & (29.3\,\%)  \\
Average flux per feature & 2.9$\times10^{16}$ Mx& 2.5$\times10^{16}$ Mx& 2.2$\times10^{16}$ Mx& 2.1$\times10^{16}$ Mx\\
\hline
Merging: total & 1.6$\times10^{20}$ Mx & 1.1$\times10^{20}$ Mx & 7.9$\times10^{19}$ Mx & 5.2$\times10^{19}$ Mx\\
(\%) & (41.5\,\%) & (38\,\%) & (34.6\,\%) & (28.8\,\%)  \\
Average flux per feature & 7.4$\times10^{16}$ Mx& 6.0$\times10^{16}$ Mx& 5.5$\times10^{16}$ Mx& 5.4$\times10^{16}$ Mx\\
\hline
Splitting-off: total & 5.8$\times10^{18}$ Mx & 1.1$\times10^{19}$ Mx  & 1.9$\times10^{19}$ Mx & 2.6$\times10^{19}$ Mx\\
(\%)  & (1.5\,\%) & (4\,\%) & (8\,\%) & (14.6\,\%) \\
Average flux per feature & 8.9$\times10^{15}$ Mx& 8.9$\times10^{15}$ Mx& 9.9$\times10^{15}$ Mx& 1.0$\times10^{16}$ Mx\\
\hline
Emerging: total & 1.9$\times10^{18}$ Mx & 1.4$\times10^{18}$ Mx & 9.4$\times10^{17}$ Mx & 5.7$\times10^{17}$ Mx\\
(\%)  & (0.5\,\%) & (0.5\,\%) & (0.4\,\%) &  (0.3\,\%) \\
Average flux per feature & 5.4$\times10^{15}$ Mx& 5.7$\times10^{15}$ Mx& 5.9$\times10^{15}$ Mx& 5.8$\times10^{15}$ Mx\\
\hline\hline
Total & 4.0$\times10^{20}$ Mx & 2.9$\times10^{20}$ Mx & 2.3$\times10^{20}$ Mx & 1.8$\times10^{20}$ Mx\\
(\%)  &  (100\,\%) & (100\,\%) & (100\,\%) & (100\,\%) \\
\hline\hline
Type of death  & 10:1 & 5:1 & 3:1 & 2:1 \\
\hline
Disappearance: total & 4.3$\times10^{19}$ Mx & 4.3$\times10^{19}$ Mx & 4.3$\times10^{19}$ Mx & 4.3$\times10^{19}$ Mx\\
(\%) & (10.3\,\%) & (13.7\,\%) & (17.3\,\%) & (21.4\,\%)\\
Average flux per feature &5.8$\times10^{15}$ Mx & 5.8$\times10^{15}$ Mx& 5.8$\times10^{15}$ Mx& 5.8$\times10^{15}$ Mx\\
\hline
Splitting: total & 1.9$\times10^{20}$ Mx & 1.3$\times10^{20}$ Mx & 8.8$\times10^{19}$ Mx & 5.7$\times10^{19}$ Mx\\
(\%) & (45.2\,\%) & (41.6\,\%) & (35\,\%) & (28\,\%)  \\
Average flux per feature & 6.7$\times10^{16}$ Mx& 5.8$\times10^{16}$ Mx& 5.0$\times10^{16}$ Mx& 4.8$\times10^{16}$ Mx\\
\hline
Merging: total & 1.6$\times10^{20}$ Mx & 1$\times10^{20}$ Mx & 7.3$\times10^{19}$ Mx & 4.8$\times10^{19}$ Mx \\
(\%) & (38.4\,\%) & (33.4\,\%) & (29.1\,\%) & (23.6\,\%)  \\
Average flux per feature & 3.1$\times10^{16}$ Mx& 2.5$\times10^{16}$ Mx& 2.3$\times10^{16}$ Mx& 2.3$\times10^{16}$ Mx\\
\hline
Merging-into: total & 1.8$\times10^{19}$ Mx& 2.8$\times10^{19}$ Mx& 3.9$\times10^{19}$ Mx& 4.8$\times10^{19}$ Mx\\
(\%) & (4.4\,\%) & (9\,\%) & (15.7\,\%) & (23.4\,\%)  \\
Average flux per feature & 2.9$\times10^{16}$ Mx& 2.3$\times10^{16}$ Mx& 2.1$\times10^{16}$ Mx& 1.9$\times10^{16}$ Mx\\
\hline
Cancelling: total & 7.3$\times10^{18}$ Mx & 7.3$\times10^{18}$ Mx & 7.3$\times10^{18}$ Mx & 7.3$\times10^{18}$ Mx\\
(\%) & (1.7\,\%) & (2.3\,\%) & (2.9\,\%) & (3.6\,\%)  \\
Average flux per feature & 4.7$\times10^{15}$ Mx& 4.7$\times10^{15}$ Mx& 4.7$\times10^{15}$ Mx& 4.7$\times10^{15}$ Mx\\
\hline\hline
Total & 4.2$\times10^{20}$ Mx & 3.1$\times10^{20}$ Mx & 2.5$\times10^{20}$ Mx & 2.$\times10^{20}$ Mx\\
(\%) & (100\,\%) & (100\,\%) & (100\,\%) & (100\,\%)  \\
\hline\hline
\end{tabular}
\end{center}
\end{table*}

\begin{table*}
\begin{center}
\caption{Normalized, maximum flux distribution of the features involved in the
various types of birth and death processes.
}
\vspace{0.6cm}
\label{table_4}
\begin{tabular}{crrrr}
\hline\hline
 & \multicolumn{4}{c}{Area-ratio criteria} \\
\hline\hline
Type of birth & 10:1 & 5:1 & 3:1 & 2:1 \\
\hline
Appearance & 19.1\,\% & 25.1\,\% & 32\,\% & 39.5\,\% \\
\hline
Splitting & 41.8\,\%  & 36\,\% & 29.2\,\% & 22\,\%\\
\hline
Merging  & 36.5\,\% & 33\,\% & 29.1\,\% & 23\,\% \\
\hline
Splitting-off & 1.6\,\% & 5.0\,\% & 9.0\,\% & 15\,\% \\
\hline
Emerging  & 1.0\,\% & 0.9\,\% & 0.7\,\% & 0.5\,\% \\
\hline\hline
Total  &  100\,\% & 100\,\% & 100\,\% & 100\,\% \\
\hline\hline
Type of death  & 10:1 & 5:1 & 3:1 & 2:1 \\
\hline
Disappearance  & 16.2\,\% & 20.8\,\% & 25.3\,\% & 30.9\,\% \\
\hline
Splitting  & 42.3\,\% & 38.4\,\% & 32.2\,\% & 25\,\% \\
\hline
Merging  & 34.7\,\% & 29.4\,\% & 24.6\,\% & 19.5\,\% \\
\hline
Merging-into & 3.9\,\% & 7.8\,\% & 13.3\,\% & 19\,\% \\
\hline
Cancelling  & 2.9\,\% & 3.6\,\% &  4.6\,\% & 5.6 \,\% \\
\hline\hline
Total & 100\,\% & 100\,\% & 100\,\% & 100\,\%  \\
\hline\hline
\end{tabular}
\end{center}
\end{table*}

\begin{table*}
\begin{center}
\caption{Flux distribution in different types of emergence and cancellation 
events (see Sect.~\ref{bipol}).}
\vspace{0.6cm}
\label{table_5}
\begin{tabular}{crrrr}
 & \multicolumn{4}{c}{Emergence events} \\
\hline
Area-ratio & 10:1 & 5:1 & 3:1 & 2:1 \\
\hline
Fractional instantaneous flux & \multicolumn{4}{c}{} \\
\hline
Time-symmetric emergence & 3\,\% & 4\,\% & 6\,\% & 10\,\% \\
Time-asymmetric emergence & 97\,\% & 96\,\% & 94\,\% & 90 \,\% \\
\hline
Fractional maximum flux & \multicolumn{4}{c}{} \\
\hline
Time-symmetric emergence & 4\,\% &  5\,\% &  9\,\% & 14\,\% \\
Time-asymmetric emergence & 96\,\% & 95\,\% & 91\,\% & 86 \,\% \\
\hline
 & \multicolumn{4}{c}{Cancellation events} \\
\hline
Area-ratio & 10:1 & 5:1 & 3:1 & 2:1 \\
\hline
Fractional instantaneous flux & \multicolumn{4}{c}{} \\
\hline
Complete cancellation & 2\,\% & 2\,\% & 2\,\% & 2\,\% \\
Semi and partial cancellation & 98\,\% & 98\,\% & 98\,\% & 98\,\% \\
\hline
Fractional maximum flux & \multicolumn{4}{c}{} \\
\hline
Complete cancellation & 3\,\% & 2\,\% & 2\,\% &  2\,\% \\
Semi and partial cancellation & 97\,\% & 98\,\% & 98\,\% & 98\,\% \\
\hline
\end{tabular}
\end{center}
\end{table*}

\begin{table*}
\begin{center}
\caption{Comparison of our results with previous studies (see Sect.~\ref{disc} for details).}
\vspace{0.6cm}
\label{table_6}
\begin{tabular}{crrrr}
\hline
Birth events & Previous studies & Present study & Present study \\
 &  & (fractional instantaneous flux) & (fractional maximum flux) \\
\hline
Appearance &  12\,\% \citep[][]{lambetal2008} & 12\,\% & 19.1\,\% \\
\hline
Fragmentation & 76\,\% \citep[][]{lambetal2008} & 44.5\,\% splitting & 41.8\,\% splitting \\
(Splitting/Merging)                           &  & 41.5\,\% merging   & 36.5\,\% merging \\
\hline
Emergence & 1\,\% \citep[][]{lambetal2008}  & 1\,\% & 2\,\% \\
\hline
Death events & Previous studies & Present study & Present study \\
 &  & (instantaneous flux) & (maximum flux) \\
\hline
Disappearance & 83\,\% \citep[][]{lambetal2013} & 86\,\% & 85\,\%\\
\hline
Cancellation & 12\,\% \citep[][]{lambetal2013} & 14\,\% & 15\,\% \\
 &  &  &  \\
\hline
\end{tabular}
\end{center}
\end{table*}

\begin{table*}
\begin{center}
\caption{Parameters of fits to lifetime distributions.}
\vspace{0.6cm}
\label{table_7}
\begin{tabular}{crrrr}
\hline
 & \multicolumn{4}{c}{Area-ratio criteria} \\
\hline
 & 10:1 & 5:1 & 3:1 & 2:1 \\
\hline
Power law index  & -3.7 & -3.7 & -3.3 & -3.2 \\
$\chi^2$ of the fit & 7.1 & 5.9 & 3.5 & 3.5\\
\hline
Exponential-fit index & -0.7 & -0.6 & -0.5 & -0.5 \\
$\chi^2$ of the fit & 3.6 & 4.9 & 3.8 & 4.4\\
\hline
\end{tabular}
\end{center}
\end{table*}
\vspace{1cm}
\begin{acknowledgements}
L.S.A. would like to thank the Alexander von Humboldt foundation for the 
fellowship that supported this project at MPS, G\"ottingen, Germany. 
The authors would like to thank Dr. Andreas Lagg for useful discussions.
\end{acknowledgements}

\appendix
\section{Quantitative definitions of splitting and merging events}
\label{appendixa}

Below we provide precise quantitative definitions of splitting and merging 
events. 

Let $N_a$ denote the number of features at time $t_2$ that spatially overlap with 
a feature labeled as $a$ at $t_1$. Let $N_b$ denote the number of features at
$t_1$ that spatially overlap with a feature labeled as $b$ at $t_2$. Let 
$b_i, i=1,2,...,N_a$ denote the features that spatially overlap with $a$, and 
$a_i, i=1,2,...,N_b$ denote the features that spatially overlap with $b$. 
Let both $b_i$ and $a_i$ be ordered, in decreasing order of their areas. 
Let $n_s:1$ denote the general area-ratio criterion, with $n_s=2,3,5$.\\

\noindent 
$\bullet$ Criteria for an event to be classified as a splitting event, a disappearance, 
or not to be considered as a death/birth event at all :  \\

\noindent
$N_a > 1$ : If $n_s \times$ area($b_2$) > area($b_1$) then 
$a$ splits into $b_i, i=1,2,...,N_a$. If $n_s \times$ 
area($b_2$) < area($b_1$) then $a$ continues to live as $b_1$
and the remaining features $b_i, i=2,...,N_a$ are considered to be
newly born at $t_2$ by splitting off from $a$. \\

\noindent
$N_a = 1$ : In this case, $a$ spatially overlaps with only one feature $b_1$
and therefore $a$ continues to live as $b_1$ at $t_2$.\\

\noindent
$N_a = 0$ : Here, $a$ does not spatially overlap with any of the
features at $t_2$ and hence it simply dies at $t_1$. \\ 

\noindent 
$\bullet$ Criteria for an event to be classified as a merging event, an appearance, or not 
to be considered as a death/birth event:  \\

\noindent
$N_b > 1$ : If $n_s \times$ area($a_2$) > area($a_1$) then 
$a_i, i=1,2,...,N_b$ merge together to form the feature $b$. If 
$n_s \times$ area($a_2$) < area($a_1$) then $a_1$ continues to 
live as $b$ and the remaining features $a_i, i=2,...,N_b$ are considered to 
be dead at $t_1$ by merging into $b$. \\

\noindent
$N_b = 1$ : Here, $b$ spatially overlaps with only one feature $a_1$
and therefore $a_1$ at $t_1$ continues to live as $b$ at $t_2$.\\

\noindent
$N_b = 0$ : Here, $b$ does not spatially overlap with any of the
features at $t_1$ and hence it is simply born at $t_2$. \\ 

\section{Simultaneous events and ambiguities}
\label{appendixb}
From one time step to the next, say, $t_1$ to $t_2$, either a feature may 
continue to live or it may die, not continuing to live on to $t_2$, i.e. 
the feature dies at $t_1$. Clearly, a feature cannot live on and die at 
the same time. However, in some cases described below,
when multiple events happen simultaneously, ambiguities regarding
life and death of a given feature can arise. To resolve such ambiguities, first 
we define priorities for different events. \\

\noindent
(1) Primary, i.e. highest priority, events : Death of parent features due to splitting or 
death of parent features due to merging. \\

\noindent
(2) Secondary events : Feature with largest area continues to live through splitting 
or merging event, because area-ratio is not met. \\

\noindent
(3) Tertiary, i.e. lowest priority, events : Death of features with smaller area 
when feature with largest area lives on. This happens when the area-ratio criterion is not 
met for merging features. \\

A given feature $a$ at a given time $t_1$ can undergo one of the following 
events independently.\\

\noindent
(i) $E_1$ : Split into child features, say into $b_1$ and $b_2$, 
while satisfying an imposed area-ratio condition.\\

\noindent
(ii) $E_2$ : Merge with another parent feature $a_1$ while satisfying an imposed 
area-ratio condition, to form a new feature $b$.\\

\noindent
(iii) $E_3$ : Split into multiple features, say $b_3$ and $b_4$, 
but not satisfying the imposed area-ratio 
condition. So in effect the feature $a$ lives on as the feature with largest 
area among its children, i.e. among $b_3$ and $b_4$.\\

\noindent
(iv) $E_4$ : Merge with another feature, say $a_2$ not satisfying the imposed 
area-ratio condition, to form a child feature, say $b_5$.
Then the feature with largest
area among $a$ and $a_2$ lives on as $b_5$.\\

\begin{figure}
\includegraphics[scale=0.45]{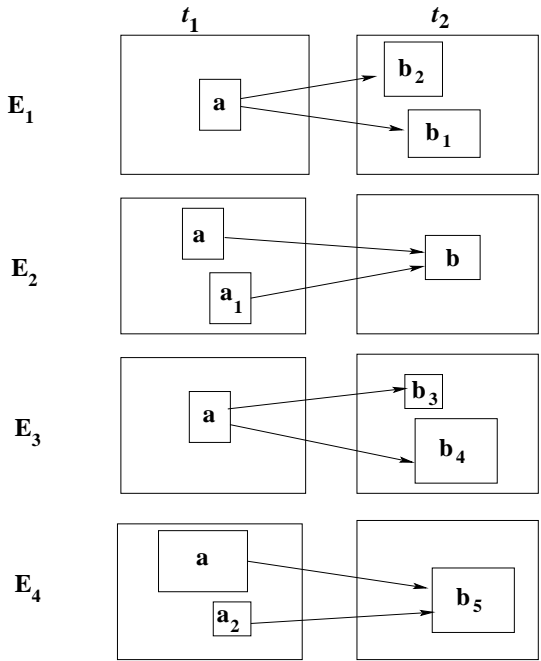}
\caption{Events $E_i$, $i=1,2,3,4$ are shown with the
conditions that area$(b_4)$ > area$(b_3)$ in $E_3$ and 
area$(a)$ > area$(a_2)$ in $E_4$.}
\label{fig-ex}
\end{figure}

In Fig.~\ref{fig-ex} a cartoon diagram describes these events 
visually, with the condition area$(b_4)$ > area$(b_3)$ in $E_3$ and 
area$(a)$ > area$(a_2)$ in $E_4$. \\

Among the above four events, we can form six non-empty sets as 
$E_i \cap E_j$, $i,j=1,2,3,4$, $i \ne j$ 
(note here that for any $i \ne j$, $E_i \cap E_j=E_j \cap E_i$). 
Although $E_1$ and $E_3$ can happen simultaneously, they can be treated 
as a single splitting event with multiple child features, which can be 
handled unambiguously. Similarly, $E_2$ and $E_4$ happening simultaneously 
can be handled as a single merging event, without any ambiguity.
Therefore we do not discuss these straightforward cases here.
We describe each of the other 4 possibilities in the following 
and the respective solutions for the ambiguities. 
In Fig.~\ref{fig-ex} we show these events in a cartoon diagram. 
Always left panels represent $t=t_1$ and the right panels $t=t_2$.
\\

\noindent
1. $E_1 \cap E_2 \ne \emptyset$ : In this case, in the same time step $a$ 
splits into two ($E_1$) and merges with another magnetic feature ($E_2$). 
Event $E_1$ leads to the death of the feature $a$, and event $E_2$ leads 
to the death of features $a$ and $a_1$,
and together these 2 events lead to the birth of features $b_1$, $b_2$ and
$b$ at $t_2$. Note that either of $b_1$, $b_2$ could be 
the same feature as $b$. Consequently, this simultaneous event does not
lead to any ambiguity.

\noindent
2. $E_1 \cap E_4 \ne \emptyset$ : In this case, $E_1$ leads to the death of
$a$ and $E_4$ leads to a continued life of $a$ or $a_2$ as $b_5$. 
Here we have two sub cases as follows. \\

\noindent
(i) area$(a)$ > area$(a_2)$ : In this case, $E_4$ leads to a 
continued life of $a$, which will live as $b_5$ at $t_2$. 
This leads to an ambiguity of life and death of $a$. Since 
$E_1$ is a primary event, it is given the higher priority 
over the secondary event $E_4$. Thus, $a$ dies as a result of $E_1$
and $a_2$ also dies because it is the smaller counter part in the merging 
event. \\

\noindent
(ii) area$(a_2)$ > area$(a)$ : In this case, $E_4$ leads to a 
continued life of $a_2$ which will live as $b_5$ at $t_2$. There is no 
ambiguity in this case. \\

\noindent
3. $E_2 \cap E_3 \ne \emptyset$ : In this case, $E_2$ leads to the death
of $a$ and $a_1$ and $E_3$ leads to a continued life of $a$ which would live on 
as $b_3$ or $b_4$. Therefore there is an ambiguity of whether $a$ lives or dies. 
Since $E_2$ is given a higher priority over $E_3$, $a$ and $a_1$ die as a
result of $E_2$, and $b_3$ and $b_4$ are considered to be newly born at $t_2$.\\

\noindent
4. $E_3 \cap E_4 \ne \emptyset$ : In this case, $E_3$ leads to a continued 
life of $a$ which will live as $b_3$ or $b_4$ and $E_4$ leads to a continued 
life of $a$ or $a_2$ as $b_5$. Note that for this simultaneous event to occur, 
we have named the features at $t_2$ such that area$(b_5)$ is smaller than 
area$(b_3)$ and area$(b_4)$. Here we have four sub cases similar to cases 3 
and 5 above. This is a complex situation because both $E_3$ and $E_4$ are not
primary events. Therefore let us consider the four sub-cases in detail. \\

\noindent
(i) area$(b_3)$ > area$(b_4)$ and area$(a)$ > area$(a_2)$  : 
Here $E_3$ leads to a continued life of $a$ as $b_3$ at $t_2$. $E_4$ leads
to a continued life of $a$ as $b_5$. The smaller counterpart of the merging
event $a_2$ will simply die. This case leads to an ambiguity between whether 
$a$ continues to live as $b_3$ or as $b_5$. Since $a$ spatially overlaps with $b_3$, $b_4$ and $b_5$,
actually $b_5$ can also be considered to be a child of the splitting of $a$. Since we have
named the features at $t_2$ such that area$(b_5)$ < area$(b_3)$ and 
area$(b_5)$ < area$(b_4)$, we let $a$ live as the feature with largest
area at $t_2$, namely, $b_3$. The features $b_4$ and $b_5$ are considered to 
be newly born.\\

\noindent
(ii) area$(b_4)$ > area$(b_3)$ and area$(a)$ > area$(a_2)$ : 
This case is the same as case (i) above with the roles of $b_3$ and $b_4$
interchanged. Here $a$ lives on as $b_4$.\\

\noindent
(iii) area$(b_3)$ > area$(b_4)$ and area$(a_2)$ > area$(a)$ : 
Here $E_3$ leads to a continued life of $a$ as $b_3$ at $t_2$. $E_4$ leads
to a continued life of $a_2$ as $b_5$. The smaller counterpart of the merging
event that dies in this case is $a$. This leads to an ambiguity between the life 
and death of $a$. \\

\noindent
Since $E_3$ leads to the life of a bigger counterpart of the 
splitting event and $E_4$ leads to the death of 
smaller counter part of the merging event. Since $E_3$ is a 
secondary event, while $E_4$ is a tertiary event, $E_3$ is
given the higher priority over $E_4$ and hence $a$ continues to live as
$b_3$. \\

\noindent
(iv) area$(b_4)$ > area$(b_3)$ and area$(a_2)$ > area$(a)$ : Here $a_2$ continues to live 
as $b_5$ at $t_2$. \\

\noindent
This case is the same as case (iii) above with the roles of 
$b_3$ and $b_4$ interchanged. \\

We remark here that two similar events can also happen simultaneously (e.g.,
two events of type $E_3$). The disambiguations employed for these cases are similar
to those discussed above. 

\section{Flow-diagrams for same polarity and 
opposite polarity interactions}
\label{appendixc}

\begin{figure*}
\includegraphics[scale=0.45]{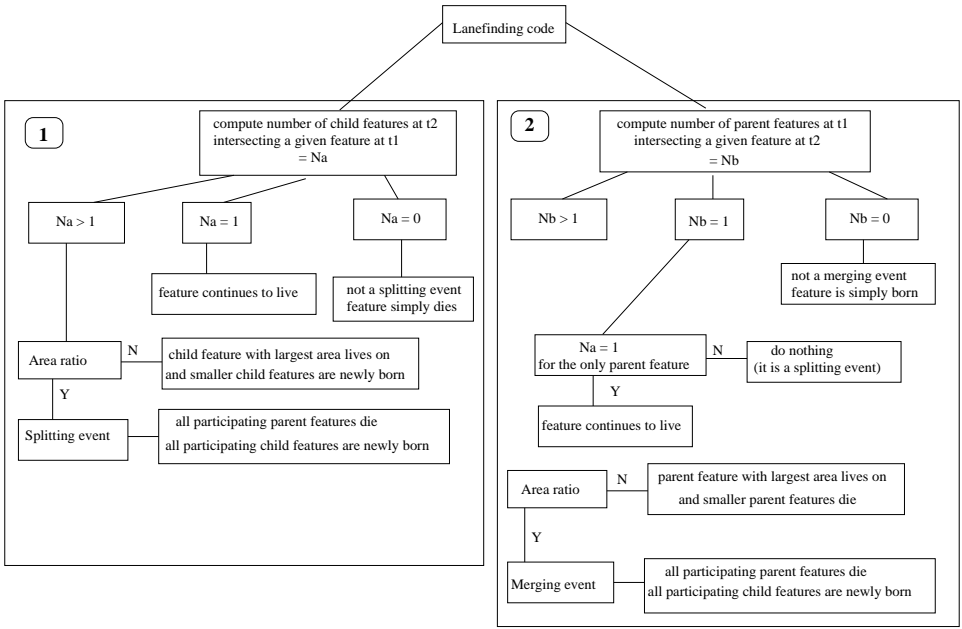}
\caption{Flowchart describing the code structure for unipolar interactions.}
\label{fig-f1}
\end{figure*}
\begin{figure*}
\includegraphics[scale=0.45]{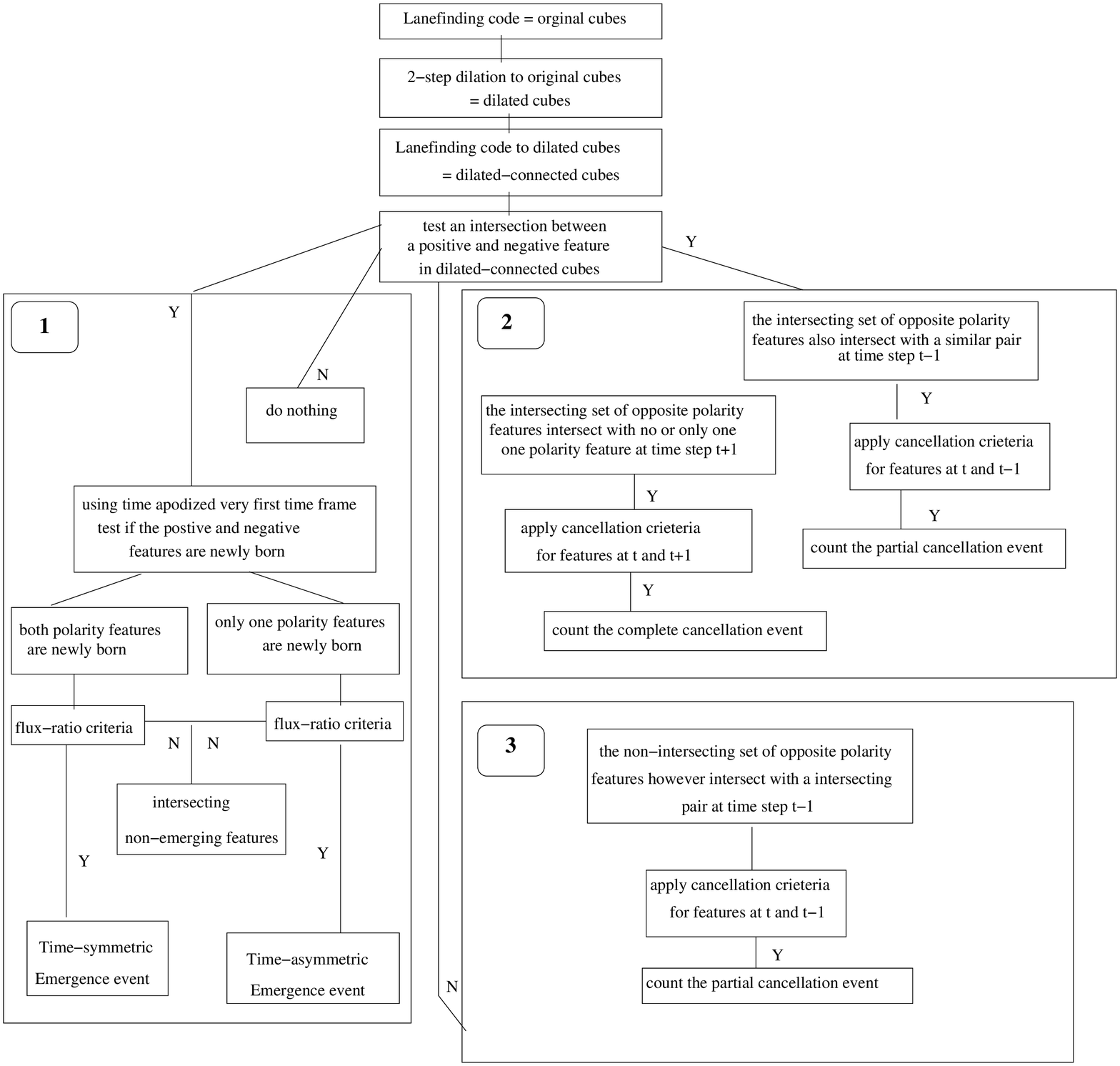}
\caption{Flowchart describing the code structure for bi-polar interactions.}
\label{fig-f2}
\end{figure*}
\subsection{Description of algorithm}
In this section we briefly give the computing algorithm for 
studies involving interactions between same-polarity
features. The terms used here are defined in Section~\ref{def}. 
A flowchart describing the corresponding computer code is given in Fig.~\ref{fig-f1}. \\

\noindent
Step 1: A lane-finding code is used to detect the features \citep[see][]{hirzbergeretal99}.\\

\noindent
Step 2: The features at two successive time steps are checked for a spatial overlap. 
The number of features at time step 
$t_2$ which spatially overlaps with each of the features at $t_1$ are counted.\\

\noindent
Step 3: If the parent feature $a_1$ at $t_1$ spatially overlaps with more than one 
child feature at $t_2$, then the areas of all these child features at $t_2$ 
are calculated. An area-ratio criterion is checked between the child 
features. We consider four cases, namely, 2:1, 3:1, 5:1 or 10:1.\\

\noindent
Step 4: If the case-specific 
area-ratio criterion is satisfied between the two largest 
child features, then it is counted as a splitting event.\\

\noindent
Step 5: If the area-ratio criterion is not satisfied, then the parent feature
is considered to live on at time $t_2$ without splitting, in the form of the 
largest child feature. All the smaller child features are 
counted as newly born features at $t_2$, classified under splitting-off birth 
events.\\

\noindent
Step 6: If the parent feature $a_1$ at $t_1$ spatially overlaps with only one child 
feature at $t_2$, then either it is considered to live on without
any interaction at $t_2$, or it could be participating in a merging event. 
In the second case, the decision made at this stage will be overridden by
the decisions taken at the stage where the merging events are studied in this
sequence.\\

\noindent
Step 7: If the parent feature $a_1$ at $t_1$ spatially overlaps with no child features
at $t_2$, then it is considered to be dead without any interaction 
(disappearing feature).\\

\noindent
Step 8: The number of features which stay up to the last time step of the 
observations are counted and are considered to live on (form of death remains undefined).\\

\noindent
Step 9: At the same time step $t_1$, the number of features that spatially overlap  
with each of the features at $t_2$ are counted.\\

\noindent
Step 10: If more than one parent features at $t_1$ spatially overlaps with a given feature
$a_2$ at $t_2$, then the areas of all these parent features at $t_1$ 
are calculated. An area-ratio criterion is checked between the parent 
features. We consider four cases, namely, 2:1, 3:1, 5:1 or 10:1.\\

\noindent
Step 11: If the case-specific area-ratio criterion is satisfied between the 
two largest parent features, then it is counted as a merging event.\\

\noindent
Step 12: An additional check is made for the presence of simultaneous splitting 
and merging events (see Appendix~\ref{appendixb} for more details).\\

\noindent
Step 13: If the area-ratio criterion is not satisfied, then the largest parent feature
is considered to live on at $t_2$ without merging. All the smaller parent features are 
counted as features that die at $t_1$ classified under merging-into death events.\\

\noindent
Step 14: If the child feature $a_2$ at $t_2$ spatially overlaps with only one parent
feature at $t_1$, and if that one parent feature, say, $a_1$ also spatially 
overlaps with only one feature $a_2$ at $t_2$, then $a_1$ is considered to live on 
without any interaction at $t_2$.\\

\noindent
Step 15: If the child feature $a_2$ at $t_2$ spatially overlaps with no parent features
at $t_1$, then it is a newly born feature and no special considerations
are needed for this case (appearing features).\\

\subsection{Description of algorithm : Mixed polarity interaction}
In this section we briefly describe the computing algorithm for  
studies involving interactions between opposite polarity
features. A flowchart describing the corresponding computer code is given in 
Fig.~\ref{fig-f2}. \\

\noindent
Step 1: A lane-finding code is used to detect features with both the
polarities. The resulting data sets are stored as two cubes, one for positive 
and the other for negative features. These cubes are referred to as the 
original cubes.\\

\noindent
Step 2: A two-step dilation is applied to the original cubes and the resulting
data sets are stored as two dilated cubes.\\

\noindent
Step 3: The dilated cubes are again used as input cubes in the lane finding 
code to detect the features which are connected after the dilation (meaning, 
spatially overlapping after dilation). The 
resulting data sets are again stored as two cubes. These are referred to as 
dilated-connected cubes.\\

\noindent
Step 4: At time step $t_1$, first a spatial overlap of one or more positive 
features with one or more negative features are detected using the dilated-connected 
cubes (see step 3 above for definition). Then we apply a flux-ratio 
criterion between the sets of positive and negative features. We consider four cases, namely, 
10:1, 5:1, 3:1 and 2:1. I.e. if the ratio of total flux contained in the 
interacting positive polarity features to that in the interacting negative 
polarity features satisfies the imposed flux ratio criterion, then it is 
detected to be an emergence event.\\

\noindent
Step 5: Further, we check if the features participating in the emergence 
event are time-symmetric or time-asymmetric (see Section~\ref{def}).\\

\noindent
Step 6: If at any time step, a spatial overlap of two sets of opposite polarity features 
leads to disappearance of both the sets of opposite polarity features, then we say it is a 
complete cancellation event. If the spatial overlap  of two sets of opposite
polarity features is found to satisfy the condition for cancellation 
according to Equation~(\ref{eq-ca}), 
leading to disappearance of features of only one polarity, then the event
is considered to be a semi cancellation event.\\
 
\noindent
Step 7: If at time step, a spatial overlap of two sets of opposite
polarity features is found to satisfy the condition for cancellation
as described in Equation~(\ref{eq-ca}), leading
to survival of both the polarity features, then the
event is identified as a partial cancellation event.\\

\end{document}